\documentclass[]{pasj01}
\usepackage{url}
\newenvironment{trueauthors}{\section*{Author contributions}\fontsize{8}{11}\selectfont}{\par}
\usepackage{multirow}

\begin{document} 
\Received{2017/09/07}
\Accepted{2017/11/16}
%\Published{yyyy/mm/dd}

\title{Hitomi Observation of Radio Galaxy NGC~1275: 
The First X-ray Microcalorimeter Spectroscopy of Fe-K$\alpha$ Line Emission from an  Active Galactic Nucleus
\thanks{The corresponding authors are Hirofumi Noda, Yasushi Fukazawa, Frederick S. Porter, Laura W. Brenneman, Koichi Hagino, Taiki Kawamuro, Shinya Nakashima, Christopher S. Reynolds, and Takaaki Tanaka.
}
}
\author{Hitomi Collaboration:
Felix \textsc{Aharonian}\altaffilmark{1},
Hiroki \textsc{Akamatsu}\altaffilmark{2},
Fumie \textsc{Akimoto}\altaffilmark{3},
Steven W. \textsc{Allen}\altaffilmark{4,5,6},
Lorella \textsc{Angelini}\altaffilmark{7},
Marc \textsc{Audard}\altaffilmark{8},
Hisamitsu \textsc{Awaki}\altaffilmark{9},
Magnus \textsc{Axelsson}\altaffilmark{10},
Aya \textsc{Bamba}\altaffilmark{11,12},
Marshall W. \textsc{Bautz}\altaffilmark{13}
% ...
}

\altaffiltext{1}{Dublin Institute for Advanced Studies, 31 Fitzwilliam Place, Dublin 2, Ireland}
\altaffiltext{2}{SRON Netherlands Institute for Space Research, Sorbonnelaan 2, 3584 CA Utrecht, The Netherlands}
\altaffiltext{3}{Department of Physics, Nagoya University, Furo-cho, Chikusa-ku, Nagoya, Aichi 464-8602}
\altaffiltext{4}{Kavli Institute for Particle Astrophysics and Cosmology, Stanford University, 452 Lomita Mall, Stanford, CA 94305, USA}
\altaffiltext{5}{Department of Physics, Stanford University, 382 Via Pueblo Mall, Stanford, CA 94305, USA}
\altaffiltext{6}{SLAC National Accelerator Laboratory, 2575 Sand Hill Road, Menlo Park, CA 94025, USA}
\altaffiltext{7}{NASA, Goddard Space Flight Center, 8800 Greenbelt Road, Greenbelt, MD 20771, USA}
\altaffiltext{8}{Department of Astronomy, University of Geneva, ch. d'\'Ecogia 16, CH-1290 Versoix, Switzerland}
\altaffiltext{9}{Department of Physics, Ehime University, Bunkyo-cho, Matsuyama, Ehime 790-8577}
\altaffiltext{10}{Department of Physics and Oskar Klein Center, Stockholm University, 106 91 Stockholm, Sweden}
\altaffiltext{11}{Department of Physics, The University of Tokyo, 7-3-1 Hongo, Bunkyo-ku, Tokyo 113-0033}
\altaffiltext{12}{Research Center for the Early Universe, School of Science, The University of Tokyo, 7-3-1 Hongo, Bunkyo-ku, Tokyo 113-0033}
\altaffiltext{13}{Kavli Institute for Astrophysics and Space Research, Massachusetts Institute of Technology, 77 Massachusetts Avenue, Cambridge, MA 02139, USA}
% ...
%% \altaffiltext{77}{Department of Physics, Faculty of Science and Engineering, Konan University, 8-9-1 Okamoto, Kobe, Hyogo 658-8501}
%% \altaffiltext{78}{Kavli Institute for the Physics and Mathematics of the Universe (WPI), The University of Tokyo, 5-1-5 Kashiwanoha, Kashiwa, Chiba 277-8583}
%% \altaffiltext{79}{National Astronomical Observatory of Japan, 2-21-1 Osawa, Mitaka, Tokyo 181-8588}

\email{hirofumi.noda@astr.tohoku.ac.jp}

\KeyWords{galaxies: active -- galaxies: individual (NGC~1275) -- galaxies: Radio galaxy -- X-rays: galaxies --Methods: observational}

\maketitle

%----------------------------------------abstract--------------------------------------------------------
\begin{abstract}
The origin of the narrow Fe-K$\alpha$ fluorescence line at 6.4~keV from active galactic nuclei has long been under debate; some of the possible sites are the outer accretion disk, the broad line region, a molecular torus, or interstellar/intracluster media. In February--March 2016, we performed the first X-ray microcalorimeter spectroscopy with the Soft X-ray Spectrometer (SXS) onboard the Hitomi satellite of the Fanaroff-Riley type I radio galaxy NGC~1275 at the center of the Perseus cluster of galaxies. With the high energy resolution of $\sim5$~eV at 6~keV achieved by Hitomi/SXS, we detected the Fe-K$\alpha$ line with $\sim 5.4~\sigma$ significance. The velocity width is constrained to be 500--1600~km~s$^{-1}$ (FWHM for Gaussian models) at 90\% confidence.  The SXS also constrains the continuum level from the NGC~1275 nucleus up to $\sim20$~keV, giving an equivalent width $\sim 20$~eV of the 6.4~keV line. Because the velocity width is narrower than that of broad H$\alpha$ line of $\sim2750$~km~s$^{-1}$, we can exclude a large contribution to the line flux from the accretion disk and the broad line region. Furthermore, we performed pixel map analyses on the Hitomi/SXS data and image analyses on the Chandra archival data, and revealed that the Fe-K$\alpha$ line comes from a region within $\sim1.6$~kpc from the NGC~1275 core, where an active galactic nucleus emission dominates, rather than that from intracluster media. Therefore, we suggest that the source of the Fe-K$\alpha$ line from NGC~1275 is likely a low-covering fraction molecular torus or a rotating molecular disk which probably extends from a pc to hundreds pc scale in the active galactic nucleus system.
\end{abstract}
%--------------------------------------------------------------------------------------------------------------

%--------------------Introduction---------------------
\section{Introduction}\label{s1}
%---------------------------------------------------------

Fluorescent Fe-K$\alpha$ emission is the most ubiquitous atomic feature in the X-ray spectrum of non-blazar type active galactic nuclei (AGN; e.g. \cite{1990Natur.344..132P, 1994MNRAS.268..405N}). The line is produced by the irradiation of low- and intermediate-ionization circumnuclear gas by the hard X-ray continuum emitted from the corona of the accreting supermassive black hole (SMBH). Thus, it provides a powerful probe of the circumnuclear environment of an AGN. 

There are at least three observable Fe-K emission components.  Irradiation of the inner accretion disk gives a relativistically-broadened line (\cite{1989MNRAS.238..729F, 1995Natur.375..659T}) that can be used to constrain the inclination of the inner disk and the black hole spin (\cite{1996MNRAS.279..837I, 2011ApJ...736..103B}). The much more distant interstellar medium and disk winds can be photo-ionized and produce narrow ionized (Fe XXV and Fe XXVI) iron lines (\cite{2003ApJ...596...85Y}). Finally, most AGN display a relatively narrow ($<3000$~km~s$^{-1}$) and neutral Fe-K$\alpha$ line at 6.4~keV. The origin of this line is unclear, with candidates being the outer edge of the accretion disk, the broad line region (BLR), the dusty torus of unified AGN schemes (\cite{1993ARA&A..31..473A, 1995PASP..107..803U}), or the larger scale reservoir from which the AGN is fueled.

%\bibitem[Pounds et al.(1990)]{1990Natur.344..132P} Pounds, K.~A., Nandra, K., Stewart, G.~C., George, I.~M., \& Fabian, A.~C.\ 1990, \nat, 344, 132
%\bibitem[Nandra \& Pounds(1994)]{1994MNRAS.268..405N} Nandra, K., \& Pounds, K.~A.\ 1994, \mnras, 268, 405 
%\bibitem[Fabian et al.(1989)]{1989MNRAS.238..729F} Fabian, A.~C., Rees, M.~J., Stella, L., \& White, N.~E.\ 1989, \mnras, 238, 729 
%\bibitem[Iwasawa et al.(1996)]{1996MNRAS.279..837I} Iwasawa, K., Fabian, A.~C., Mushotzky, R.~F., et al.\ 1996, \mnras, 279, 837 
%\bibitem[Tanaka et al.(1995)]{1995Natur.375..659T} Tanaka, Y., Nandra, K., Fabian, A.~C., et al.\ 1995, \nat, 375, 659 
%\bibitem[Brenneman et al.(2011)]{2011ApJ...736..103B} Brenneman, L.~W., Reynolds, C.~S., Nowak, M.~A., et al.\ 2011, \apj, 736, 103 
%\bibitem[Yaqoob et al.(2003)]{2003ApJ...596...85Y} Yaqoob, T., George, I.~M., Kallman, T.~R., et al.\ 2003, \apj, 596, 85
%\bibitem[Antonucci(1993)]{1993ARA&A..31..473A} Antonucci, R.\ 1993, \araa, 31, 473 
%\bibitem[Urry \& Padovani(1995)]{1995PASP..107..803U} Urry, C.~M., \& Padovani, P.\ 1995, \pasp, 107, 803

To date, most studies of the narrow Fe-K$\alpha$ line have focused on X-ray bright Seyfert galaxies utilizing both dispersive high-resolution spectrometers such as the Chandra/High-Energy Transmission Grating (HETG; \cite{2000ApJ...539L..41C}) and non-dispersive moderate-resolution spectrometers such as Chandra/Advanced CCD Imaging Spectrometer (ACIS; \cite{2003SPIE.4851...28G}) and XMM-Newton/European Photon Imaging Camera (EPIC; \cite{2001A&A...365L..27T}; \cite{2001A&A...365L..18S}).  Using HEG data, Shu et al. (2010; 2011) reported that the narrow Fe-K$\alpha$ line from typical Seyfert galaxies exhibits velocity widths greater than $\sim 2000$ km s$^{-1}$ (FWHM). Subsequently, \citet{2015ApJ...812..113G} and \citet{2015ApJ...802...98M} compared object-by-object velocity widths of the Fe-K$\alpha$ line, the broad Balmer lines of hydrogen and those corresponding to dust sublimation radii, and found that the Fe-K$\alpha$ source lies between the BLR and dust sublimation radius. In contrast, the origin of the Fe-K$\alpha$ line from radio-loud AGN remains unclear in spite of observations with these high-performance X-ray spectrometers, principally due to their relatively low equivalent widths and fluxes (see \cite{2011ApJ...727...19F}).  We need the leap in the sensitivity and resolution provided by the Soft X-ray Spectrometer (SXS; \cite{2016SPIE.9905E..0VK}) onboard the Hitomi satellite (\cite{2016SPIE.9905E..0UT}) to establish an picture for the circumnuclear distribution of gas in AGN via studies of the Fe-K$\alpha$ line. 

%\bibitem[Fukazawa et al.(2011)]{2011ApJ...727...19F} Fukazawa, Y., Hiragi, K., Mizuno, M., et al.\ 2011, \apj, 727, 19 
%\bibitem[Fukazawa et al.(2016)]{2016ApJ...821...15F} Fukazawa, Y., Furui, S., Hayashi, K., et al.\ 2016, \apj, 821, 15 
%\bibitem[Shu et al.(2010)]{2010ApJS..187..581S} Shu, X.~W., Yaqoob, T., \& Wang, J.~X.\ 2010, \apjs, 187, 581 
%\bibitem[Shu et al.(2011)]{2011ApJ...738..147S} Shu, X.~W., Yaqoob, T., \& Wang, J.~X.\ 2011, \apj, 738, 147 
%\bibitem[Canizares et al.(2000)]{2000ApJ...539L..41C} Canizares, C.~R., Huenemoerder, D.~P., Davis, D.~S., et al.\ 2000, \apjl, 539, L41 
%\bibitem[Koyama et al.(2007)]{2007PASJ...59S..23K} Koyama, K., Tsunemi, H., Dotani, T., et al.\ 2007, \pasj, 59, 23 
%\bibitem[Gandhi et al.(2015)]{2015ApJ...812..113G} Gandhi, P., H{\"o}nig, S.~F., \& Kishimoto, M.\ 2015, \apj, 812, 113 
%\bibitem[Minezaki \& Matsushita(2015)]{2015ApJ...802...98M} Minezaki, T., \& Matsushita, K.\ 2015, \apj, 802, 98 
%\bibitem[Garmire et al.(2003)]{2003SPIE.4851...28G} Garmire, G.~P., Bautz, M.~W., Ford, P.~G., Nousek, J.~A., \& Ricker, G.~R., Jr.\ 2003, \procspie, 4851, 28
%\bibitem[Turner et al.(2001)]{2001A&A...365L..27T} Turner, M.~J.~L., Abbey, A., Arnaud, M., et al.\ 2001, \aap, 365, L27 
%\bibitem[Str{\"u}der et al.(2001)]{2001A&A...365L..18S} Str{\"u}der, L., Briel, U., Dennerl, K., et al.\ 2001, \aap, 365, L18 

In this paper, we focus on the X-ray properties of the radio-loud AGN NGC~1275 which resides at the center of the Perseus cluster of galaxies ($z_*=0.017284$; \cite{Hitomi2017a}).  Chandra and XMM-Newton observations reveal signatures of interaction between the jets from NGC~1275 and the cool-core of the intracluster medium (ICM), most notably the presence of kpc-scale cavities and weak shocks or sound waves in the ICM (e.g., \cite{2006MNRAS.366..417F}). There is compelling evidence that these jet-ICM interactions in cool core galaxy clusters heat the ICM, preventing the ICM core from undergoing a cooling catastrophe.  Being the brightest example, the Perseus cluster and NGC~1275 have thus been recognized as one of the most important targets for gaining an understanding of the feedback loop that allows a central AGN to regulate a cluster core (e.g., \cite{2007ARA&A..45..117M}).  For this reason, the core of the Perseus cluster was the first science target for the SXS onboard Hitomi. This X-ray micro-calorimeter observation produced a spectrum of the ICM with unprecedented spectral resolution, finding a surprisingly quiescent turbulent velocity of $\sim 160$~km~s$^{-1}$ despite the vigorous AGN feedback \citep{2016Natur.535..117H}. This same observation gives us our first view of an AGN, NGC~1275, with an X-ray microcalorimeter --- that is the focus of this paper.

%\bibitem[Strauss et al.(1992)]{1992ApJS...83...29S} Strauss, M.~A., Huchra, J.~P., Davis, M., et al.\ 1992, \apjs, 83, 29 
%\bibitem[Fabian et al.(2006)]{2006MNRAS.366..417F} Fabian, A.~C., Sanders, J.~S., Taylor, G.~B., et al.\ 2006, \mnras, 366, 417
%\bibitem[McNamara \& Nulsen(2007)]{2007ARA&A..45..117M} McNamara, B.~R., \& Nulsen, P.~E.~J.\ 2007, \araa, 45, 117 
%\bibitem[Hitomi Collaboration et al.(2016)]{2016Natur.535..117H} Hitomi Collaboration, Aharonian, F., Akamatsu, H., et al.\ 2016, \nat, 535, 117
%\bibitem[Kelley et al.(2016)]{2016SPIE.9905E..0VK} Kelley, R.~L., Akamatsu, H., Azzarello, P., et al.\ 2016, \procspie, 9905, 99050V
%\bibitem[Takahashi et al.(2016)]{2016SPIE.9905E..0UT} Takahashi, T., Kokubun, M., Mitsuda, K., et al.\ 2016, \procspie, 9905, 99050U 

The X-ray properties of NGC~1275 have been studied for many years.  \citet{2015MNRAS.451.3061F} compiled a 40~year X-ray light curve from archival and historical multi-mission data, and compared it to the 90 GHz flux over the same time.  They found that the X-ray and radio flux were correlated.  They also found that the source was very bright in the 1970--80s, fading to a minimum in 2000-2005, and has been slowly brightening since that time.  The X-ray spectrum observed by XMM-Newton in 2001 exhibited the presence of a clear narrow Fe-K$\alpha$ line with an equivalent width (EW) of $\sim165$~eV relative to the AGN continuum \citep{2003ApJ...590..225C}.  A later XMM-Newton observation in 2006 found that the EW had decreased to 70--80~eV \citep{2013PASJ...65...30Y}, mostly due to an increase in the AGN continuum flux, while the line flux was roughly constant. However, the line width has never been successfully measured, and hence, it was not known which region mainly generates the Fe-K$\alpha$ emission in NGC~1275. Here we report the first measurements of the Fe-K$\alpha$ line width in NGC~1275, providing the such measurement from a radio-loud, central cluster AGN.  This is the first X-ray spectroscopy of an AGN with an X-ray microcalorimeter. 

%\bibitem[Fabian et al.(2015)]{2015MNRAS.451.3061F} Fabian, A.~C., Walker, S.~A., Pinto, C., Russell, H.~R., \& Edge, A.~C.\ 2015, \mnras, 451, 3061 
%\bibitem[Churazov et al.(2003)]{2003ApJ...590..225C} Churazov, E., Forman, W., Jones, C., \& B{\"o}hringer, H.\ 2003, \apj, 590, 225 
%\bibitem[Yamazaki et al.(2013)]{2013PASJ...65...30Y} Yamazaki, S., Fukazawa, Y., Sasada, M., et al.\ 2013, \pasj, 65, 30 

NGC~1275 harbors a SMBH with a mass $M_{\rm BH} \sim 3.4\times10^8~M_{\odot}$ \citep{2005MNRAS.359..755W} or $\sim 8\times10^8~M_{\odot}$ \citep{2013MNRAS.429.2315S}, and exhibits a relatively low Eddington ratio, $L_B/L_{\rm Edd} \sim 3\times 10^{-4}$ (e.g., \cite{2007ApJ...658..815S}).  The AGN has been often classified as a Fanaroff-Riley type-I (FR~I) radio galaxy \citep{1983MNRAS.204..151L}, but it is still unclear whether or not this system has a Seyfert-like BLR.  While a broad H$\alpha$ line with velocity width of $\sim 2750$~km~s$^{-1}$ (FWHM) was observed during the bright phase in 1984 \citep{1997ApJS..112..391H}, no optical broad lines were detected by HST when the source was in the faint phase in 2000 \citep{2014A&A...563A.119B}. In the infrared band, \citet{2013MNRAS.429.2315S} reported [Fe II], H$_2$, He I, and Br$\gamma$ lines with velocity widths of $\sim 380$--1000~km~s$^{-1}$ (FWHM) from a resolved rotating molecular disk which has a size of a few hundreds pc.

In the present paper, we analyze not only the Fe-K$\alpha$ line emission from NGC~1275 but also the broad-band continuum from this AGN.  The continuum studies are enabled by fact that the SXS (at the focal point of one of the Hitomi Soft X-ray Telescopes; SXT) has sensitivity up to $\sim20$~keV, an energy at which the AGN dominates over the ICM emission, and a very low background. In Section 2, we briefly describe the Hitomi observation and data reduction as well as the data from other satellites that we shall use in this investigation.  Our analyses and results are reported in Section 3, focusing firstly on the nature of the Fe-K$\alpha$ line (Section 3.1) and then on the broad-band AGN emission (Section 3.2).  We discuss the astrophysical implications in Section 4, developing the case that the Fe-K$\alpha$ line originates from the torus or the molecular disk seen by \citet{2013MNRAS.429.2315S}.  Unless otherwise stated, tables and figures use 90\% and $1\sigma$ error ranges, respectively. All abundances are quoted relative to the \citet{2009M&PSA..72.5154L} solar abundance set.  Photoelectric cross-sections for the absorption model are obtained from \citet{Balu92}.

%\bibitem[Sikora et al.(2007)]{2007ApJ...658..815S} Sikora, M., Stawarz, {\L}., \& Lasota, J.-P.\ 2007, \apj, 658, 815 
%\bibitem[Ho et al.(1997)]{1997ApJS..112..391H} Ho, L.~C., Filippenko, A.~V., Sargent, W.~L.~W., \& Peng, C.~Y.\ 1997, \apjs, 112, 391
%\bibitem[Balmaverde \& Capetti(2014)]{2014A&A...563A.119B} Balmaverde, B., \& Capetti, A.\ 2014, \aap, 563, A119 
%\bibitem[Wilman et al.(2005)]{2005MNRAS.359..755W} Wilman, R.~J., Edge, A.~C., \& Johnstone, R.~M.\ 2005, \mnras, 359, 755
%\bibitem[Strauss et al.(1992)]{1992ApJS...83...29S} Strauss, M.~A., Huchra, J.~P., Davis, M., et al.\ 1992, \apjs, 83, 29 
%\bibitem[Laing et al.(1983)]{1983MNRAS.204..151L} Laing, R.~A., Riley, J.~M., \& Longair, M.~S.\ 1983, \mnras, 204, 151
%\bibitem[Lodders et al.(2009)]{2009LanB...4B...44L} Lodders, K., Palme, H., \& Gail, H.-P.\ 2009, Landolt B{\"o}rnstein, 
%\bibitem[Anders \& Grevesse(1989)]{Ande89} Anders, E., \& Grevesse, N.\ 1989, \gca, 53, 197 
%\bibitem[Balucinska-Church \& McCammon(1992)]{Balu92} Balucinska-Church, M., \& McCammon, D.\ 1992, \apj, 400, 699 
%\bibitem[Scharw{\"a}chter et al.(2013)]{2013MNRAS.429.2315S} Scharw{\"a}chter, J., McGregor, P.~J., Dopita, M.~A., \& Beck, T.~L.\ 2013, \mnras, 429, 2315

%==================S2===================
\section{Observation and Data reduction}\label{s2}
%==================S2===================

%%%%%%%%%%%Table 1%%%%%%%%%%                                                                                                                              
\renewcommand{\arraystretch}{1}
\begin{table*}[t]
 \caption{A list of the archival data of XMM-Newton and Chandra.}
 \label{all_tbl}
 \begin{center}
  \begin{tabular}{ccccc}
   \hline\hline
    OBSID & Observation date  & Exposure (ksec) & Angular offset ($'$) & Source / BGD extraction regions\\\hline
  \multicolumn{5}{c}{XMM-Newton/PN and MOS}\\[1.5ex]
  0085110101 & 2001 January 30 & 60$^{a}$  & 0.001 & \multirow{2}{*}{$14''$-circle / $60''$--$63''$-annulus}\\
  0305780101 & 2006 January 29 & 141$^{a}$ & 0.038 \\[1.5ex]
  \hline
   \multicolumn{5}{c}{Chandra/HEG}\\[1.5ex]
  333 & 1999 October 10 & 27 & 0.011 & \multirow{2}{*}{$4''$-circle / None} \\
  428 & 2000 August 25 & 25 & 0.011 \\[1.5ex]                                                                    
   \hline
    \multicolumn{5}{c}{Chandra/ACIS}\\[1.5ex]
  % memo. nominal point of all the obs. is (RA, Dec.) = (49.9508 41.5117)  
  6139 & 2004 October 4  & 56  & 0.007 &\multirow{10}{*}{$4''$--$45''$-annulus / off-source $50''$-circle} \\
  4946 & 2004 October 6  & 24  & 0.007 \\
  4948 & 2004 October 9  & 119 & 0.007 \\
  4947 & 2004 October 11 & 30  & 0.007 \\
  4949 & 2004 October 12 & 29  & 0.007 \\
  4950 & 2004 October 12 & 97  & 0.007 \\
  4952 & 2004 October 14 & 164 & 0.007 \\
  4951 & 2004 October 17 & 96  & 0.007 \\
  4953 & 2004 October 18 & 30  & 0.007 \\
  6145 & 2004 October 19 & 85  & 0.007 \\
  6146 & 2004 October 20 & 47  & 0.007 \\[1.5ex]
  
\hline\hline
  \end{tabular}
\end{center}
{\small 
         $^{\rm a}$ The integrated total exposure for all the detectors, PN and MOS.                                                                      
}
\label{tab:OtherSat}
\end{table*}
%%%%%%%%%%%Table 1%%%%%%%%%%%%      

%--------------------------S2.1------------------------
\subsection{Hitomi}\label{s2-1}
%--------------------------S2.1------------------------

The Hitomi observations of the core of the Perseus cluster and NGC~1275 were performed on 2016 February 24--27 and March 4--7 with three different nominal pointing positions.  Here, we utilize the SXS data obtained on February 25--27 and March 4--6 (OBSID 100040020--100040050) which were derived at the same nominal position with a total on-source exposure of approximately $240$~ksec. 
%after the SXS almost reached the thermal equilibrium \citep{2016SPIE.9905E..3SF,2016SPIE.9905E..3RN}. 
The SXS successfully achieved an unprecedented in-orbit energy resolution of $\sim4.9$~eV at 6~keV in 35 calorimeter pixels which covered a field of view (FOV) of $\sim 3' \times 3'$ (\cite{2016SPIE.9905E..0WP}; \cite{2016SPIE.9905E..3UL}).  The SXT focuses X-rays onto the SXS with a half power diameter (HPD) of $1.2'$ \citep{2016SPIE.9905E..0ZO}. The in-orbit energy resolution and the HPD were better than the formal requirements of 7~eV and $1.7'$ \citep{2016SPIE.9905E..0UT}, respectively. Although energies below $\sim 2$~keV could not be observed by the SXS due to the closed gate valve on the cryostat \citep{2016SPIE.9905E..3WE}, the extension of its bandpass up to $\sim20$~keV, which was also much beyond the requirement of 12~keV \citep{2016SPIE.9905E..0UT}, proves extremely useful for this investigation of NGC~1275. The Soft X-ray Imager (SXI; \cite{2016SPIE.9905E..10T}), the Hard X-ray Imager (HXI; \cite{2016SPIE.9905E..11N}) and the Soft Gamma ray Detector (SGD; \cite{2016SPIE.9905E..13W}) were inactive during the observations of the Perseus cluster we utilize in the present paper. 

%\bibitem[Fujimoto et al.(2016)]{2016SPIE.9905E..3SF} Fujimoto, R., Takei, Y., Mitsuda, K., et al.\ 2016, \procspie, 9905, 99053S 
%\bibitem[Noda et al.(2016)]{2016SPIE.9905E..3RN} Noda, H., Mitsuda, K., Okamoto, A., et al.\ 2016, \procspie, 9905, 99053R
%\bibitem[Nakazawa et al.(2016)]{2016SPIE.9905E..11N} Nakazawa, K., Sato, G., Kokubun, M., et al.\ 2016, \procspie, 9905, 990511 
%\bibitem[Watanabe et al.(2016)]{2016SPIE.9905E..13W} Watanabe, S., Tajima, H., Fukazawa, Y., et al.\ 2016, \procspie, 9905, 990513
%\bibitem[Okajima et al.(2016)]{2016SPIE.9905E..0ZO} Okajima, T., Soong, Y., Serlemitsos, P., et al.\ 2016, \procspie, 9905, 99050Z 
%\bibitem[Porter et al.(2016)]{2016SPIE.9905E..0WP} Porter, F.~S., Boyce, K.~R., Chiao, M.~P., et al.\ 2016, \procspie, 9905, 99050W 
%\bibitem[Leutenegger et al.(2016)]{2016SPIE.9905E..3UL} Leutenegger, M.~A., Audard, M., Boyce, K.~R., et al.\ 2016, \procspie, 9905, 99053U 
%\bibitem[Eckart et al.(2016)]{2016SPIE.9905E..3WE} Eckart, M.~E., Adams, J.~S., Boyce, K.~R., et al.\ 2016, \procspie, 9905, 99053W
%\bibitem[Takahashi et al.(2016)]{2016SPIE.9905E..0UT} Takahashi, T., Kokubun, M., Mitsuda, K., et al.\ 2016, \procspie, 9905, 99050U 

The SXS cleaned datasets were reduced via the pipeline version 03.01.006.007 \citep{2016SPIE.9905E..14A}. In order to utilize the energy range above 16~keV,  we apply \texttt{sxsextend} in \texttt{HEAsoft-6.20} to the cleaned datasets, and extract events with the screenings which use the relation between energy and a rise time of every X-ray event, and with the removals of frame events using the time coincidence among pixels. Moreover, we select events with the high primary grade to use the highest resolution events. Spectra are extracted via \texttt{xselect}. After these reductions, we further apply two types of gain corrections; one is the ``$z$ correction'' and the other is the ``parabolic correction''. The $z$ correction is a linear gain correction to all the pixels in order to align the central energy of the 6.7004~keV Fe~XXV~He$\alpha$ emission line seen in the emission from the hot gas in the Perseus cluster to the redshift of NGC~1275 of $0.017284$. The parabolic correction is applied to align the central energies of emission lines in the 1.8--9.0~keV band, with the function $\Delta E_{\rm corr} = A(E_{\rm obs} - E_{\rm res})^2 + B(E_{\rm obs} - E_{\rm res})$, where $E_{\rm obs}$ and $E_{\rm res}$ are the observed energy and the redshifted Fe~XXV~He$\alpha$ energy, respectively. The two gain corrections are the same as those used in \citet{Hitomi2017b}, and are on the order of 1 eV (see Appendix in \cite{Hitomi2017b} for further details). In the present paper, we use the datasets after both $z$ correction and parabolic correction are applied, unless otherwise stated. 

In spectral analyses, the Non-X-ray Background (NXB; \cite{2016SPIE.9905E..3LK}) events are modeled by \texttt{sxsnxbgen} in \texttt{ftools}. The Redistribution Matrix File (RMF) is prepared by \texttt{sxsmkrmf} in \texttt{HEAsoft-6.20} as the extra-large matrix which can include the escape peak and electron loss continuum effects. The Ancillary-Response Files (ARFs) are prepared via \texttt{aharfgen} in \texttt{HEAsoft-6.20}. The ARF for a point-like source is created, assuming a point-like source at ($\alpha_{\rm 2000}$, $\delta_{\rm 2000}$) = (3$^h$19$^m$48$^s$.16, 41$^{\circ}$30$^{\prime}$42$^{\prime\prime}$.10). The ARF for a diffuse source is created using input a sky image from Chandra in the 1.8--9.0~keV band where the AGN region within 10~arcsec from NGC~1275 is replaced with the average adjacent brightness.

%\bibitem[Tsunemi et al.(2016)]{2016SPIE.9905E..10T} Tsunemi, H., Hayashida, K., Tsuru, T.~G., et al.\ 2016, \procspie, 9905, 990510 
%\bibitem[Soong et al.(2014)]{2014SPIE.9144E..28S} Soong, Y., Okajima, T., Serlemitsos, P.~J., et al.\ 2014, \procspie, 9144, 914428 
%\bibitem[Angelini et al.(2016)]{2016SPIE.9905E..14A} Angelini, L., Terada, Y., Loewenstein, M., et al.\ 2016, \procspie, 9905, 990514 
%\bibitem[Kilbourne et al.(2016)]{2016SPIE.9905E..3LK} Kilbourne, C.~A., Adams, J.~S., Brekosky, R.~P., et al.\ 2016, \procspie, 9905, 99053L 

%--------------------------S2.2------------------------
\subsection{Other X-ray satellites: XMM-Newton and Chandra}\label{s2-2}
%--------------------------S2.2------------------------

To discuss the spatial extent of the Fe-K$\alpha$ line source (Section~\ref{s3-1-2}), we utilize archival datasets taken with XMM-Newton and Chandra, complementary to the Hitomi data. The XMM-Newton and Chandra data are reprocessed using the \texttt{SAS-16.0.0} and \texttt{CIAO-4.8} software, respectively. 

XMM-Newton observations with nominal positions very close to NGC~1275 were conducted in January 2001 and 2006. We use the two long datasets ($> 60$~ksec) with the EPIC MOS and PN detectors (see Table~\ref{tab:OtherSat}). Exploiting the moderate angular resolution of $\sim 15''$ (HPD), we analyze the AGN spectra. 

Chandra has been intensively used to study the Perseus cluster, resulting in 31 publicly available archival datasets as of 2016. The unprecedented angular resolution ($\lesssim 1''$) enables us to precisely localize the Fe-K$\alpha$ line by discriminating between the AGN and emission from the cluster gas. We utilize all the datasets of the HETG read out onto the ACIS. The dispersed High-Energy Grating (HEG) data are analyzed, whereas the dispersed Medium-Energy Grating (MEG) data are not used because it does not cover the Fe-K$\alpha$ line energy. The other Chandra datasets were taken with ACIS in the imaging mode. Although pile-up is significant in the central region ($< 4''$; see the detail in Section~\ref{s3-1-2}), possible Fe-K$\alpha$ line flux in the outer region ($> 4''$) can be constrained. We select (1) the data sets which have an angular offset from NGC~1275 less than $0'.1$ in order to make use of high angular resolution ($0.5''$), and (2) those taken in 2004 to reduce systematic uncertainties associated with the aging of the instruments. The final datasets we use are summarized in table~\ref{tab:OtherSat}. 
%Note that the zeroth order spectra are not usuable due to the pile-up effect (see the detail in Section~\ref{s3-1-2}). 
% We extract spectra from an annular region ($4''$--$45''$) centered on NGC~1275. 

%==================S3===================
\section{Data Analysis and Results}\label{s3}
%==================S3===================

%--------------------------S3.1------------------------
\subsection{Fe-K$\alpha$ spectral and spatial analyses}\label{s3-1}
%--------------------------S3.1------------------------

We begin by examining the spectral and spatial structure of the neutral Fe-K$\alpha$ line emission.

%--------------------------S3.1.1------------------------
\subsubsection{The SXS spectrum}\label{s3-1-1}
%--------------------------S3.1.1------------------------

\begin{figure}[t]
\begin{center}
\FigureFile(75mm,75mm){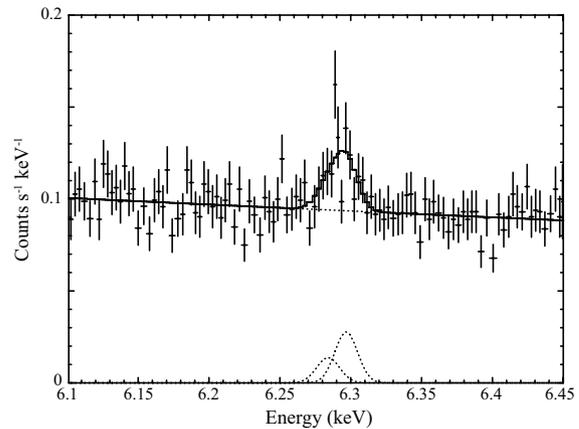}
\end{center}
\caption{The 6.1--6.45~keV SXS spectrum including the neutral Fe-K$\alpha$ line at $\sim 6.4$~keV in the rest frame. It is fitted with the model of \texttt{powerlaw + zgauss + zgauss}, which corresponds to a sum of spectral continua from the ICM and AGN, an Fe-K$\alpha_1$, and an Fe-K$\alpha_2$ emission line, respectively. }
\label{fig:Fe-Kalpha}
\end{figure}

We analyze a spectrum extracted from a field of $3 \times 3$ pixels (pixel \#0--8) of the microcalorimeter array centered on NGC~1275 (which is in pixel \#4), since  $\sim 95$\% of the AGN counts are contained in this region \citep{2016Natur.535..117H}. Figure~\ref{fig:Fe-Kalpha} shows this 9-pixels NXB-subtracted spectrum in the 6.1--6.45~keV (observed) range; a clear emission line can be seen at $\sim 6.28$~keV which corresponds to $\sim 6.40$~keV in the rest frame. Detailed spectral fitting at microcalorimeter resolution must account for the fact that the neutral Fe-K$\alpha$ line is actually a doublet with intrinsic energies 6.404~keV and 6.391~keV for Fe-K$\alpha_1$ and Fe-K$\alpha_2$, respectively.  We fit this section of the spectrum with a model consisting of a powerlaw (representing the sum of the ICM and AGN continuum) and two redshifted Gaussian lines with intrinsic energies of Fe-K$\alpha_1$ and Fe-K$\alpha_2$ (XSPEC model \texttt{powerlaw + zgauss1 + zgauss2}).  Atomic physics dictates that the intrinsic intensity of Fe-K$\alpha_1$ is twice of that of Fe-K$\alpha_2$, and so the relative normalization of the Gaussian lines are linked accordingly. Free parameters of the model are the photon index $\Gamma$ and the normalization $N_{\rm PL}$ of \texttt{powerlaw}, the redshift $z$ and overall intensity of the Fe-K$\alpha_1$/Fe-K$\alpha_2$ doublet, and the (common) velocity width $\sigma$ of the Fe-K$\alpha_1$/Fe-K$\alpha_2$ doublet.  Fitting is performed using microcalorimeter energy bins of 1~eV; we do not further bin the spectrum. We perform the spectral fitting with $C$-statistics \citep{1979ApJ...228..939C}, because of the modest count rate in each bin.

%%%%%%%%%%%%%%%%figure1%%%%%%%%%%%%%%%%%%%%%%%%
%%%%%%%%%%%%%%%%figure1%%%%%%%%%%%%%%%%%%%%%%%%

%%%%%%%%%%%%%%%%figure2%%%%%%%%%%%%%%%%%%%%%%%%
\begin{figure}[t]
\begin{center}
\FigureFile(72mm,72mm){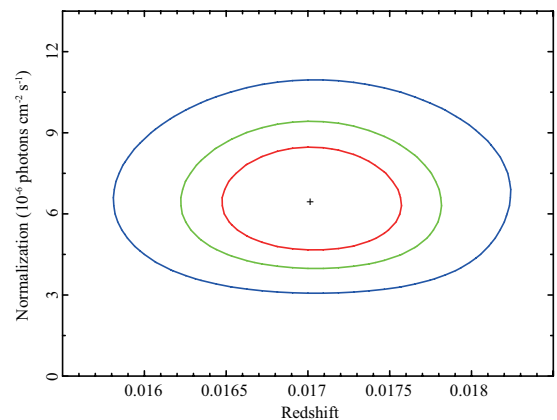}
\end{center}
\caption{Contours of confidence level of 68\% (red), 90\% (green), and 99\% (blue), between the normalization (K$\alpha_1+$K$\alpha_2$) and redshift in the Fe-K$\alpha$ line region fitting. }
\label{fig:Fe-Kalpha-cont}
\end{figure}
%%%%%%%%%%%%%%%%%figure2%%%%%%%%%%%%%%%%%%%%%%%%

%%%%%%%%%%%Table 2%%%%%%%%%%
\renewcommand{\arraystretch}{1}
\begin{table*}[t]
 \caption{Results of the fit to the 6.1--6.45~keV spectrum of the SXS on-core 9 pixels, with and without the $z$ and parabolic corrections. }
 \label{all_tbl}
 \begin{center}
  \begin{tabular}{cccc}
   \hline\hline
    Component & Parameter  & w/ $z$ and parabolic corr. & w/o $z$ and parabolic corr. \\\hline
                      	                           
  \texttt{powerlaw} & $\Gamma$
  			& $2.49^{+0.86}_{-1.22}$ & $2.55^{+0.86}_{-1.17}$\\
			
			&$N_{\rm PL}^{\rm a}$
			&$0.07^{+0.45}_{-0.06}$ & $0.08^{+0.55}_{-0.07}$\\[1.5ex] 

  \texttt{zgauss1} & $E_1$~(keV)
			& \multicolumn{2}{c}{$6.404$ (fix)}\\
			
			& $\sigma_1$~(eV)
			& $7.79^{+5.35}_{-3.37}$ & $8.93^{+5.49}_{-3.73}$\\
			
			& $z_1$
			& $0.01702^{+0.00059}_{-0.00060}$ & $0.01733^{+0.00065}_{-0.00064}$\\
			
			&$N_1^{\rm b}$
			&$4.30^{+1.48}_{-1.29}$ & $4.48^{+1.53}_{-1.34}$\\[1.5ex]                    

  \texttt{zgauss2} & $E_2$~(keV)
			& \multicolumn{2}{c}{$6.391$ (fix)}\\
			
			& $\sigma_2$~(eV)
			& \multicolumn{2}{c}{$=\sigma_1$}\\
			
			& $z_2$
			& \multicolumn{2}{c}{$=z_1$}\\
			
			&$N_2^{\rm b}$
			&\multicolumn{2}{c}{$=0.5 \times N_1$}\\[1.5ex]
			
			& EW$_1$ + EW$_2$ (eV)
			& $8.9^{+3.1}_{-2.7}$ & $9.2^{+3.2}_{-2.8}$\\[1.5ex]
			                    
    \hline                    
   $C$-statistics/d.o.f. &  & 326.2/344 & 325.5/344 \\
\hline\hline
  \end{tabular}
\end{center}
   	{\small
         $^{\rm a}$ The \texttt{powerlaw} normalization at 1 keV, in units of photons~keV$^{-1}$~cm$^{-2}$~s$^{-1}$~at 1 keV.\\
	 $^{\rm b}$ The \texttt{zgauss} normalization, in units of $10^{-6}$~photons~cm$^{-2}$~s$^{-1}$. \\         
}
\label{tab:Fe-Kalpha}
\end{table*}
%%%%%%%%%%%Table 2%%%%%%%%%%%%

This simple model provides a good description of the spectrum (with $C$-statistics/d.o.f. = 326.2/344\footnote{$C$-statistics is distributed as $\chi^2$ in the case of a large number of samples \citep{1979ApJ...228..939C}.}).  The best fitting parameter values are summarized in table~\ref{tab:Fe-Kalpha}.  Despite the fact that the sum of the EWs of the two Fe-K$\alpha$ lines (EW$_1$ + EW$_2$) is just $\sim9$~eV, they are detected with a high significance level of $\sim 5.4~\sigma$ thanks to the high energy resolution of the SXS.  This clearly illustrates the line detecting capability of X-ray microcalorimeters.  After including the effects of the instrumental broadening in the modeling, the Fe-K$\alpha$ lines have a resolved velocity width with the Gaussian sigma of $4.4$--$13.1$~eV (90\% confidence range), corresponding to a velocity width of $500$--1500~km~s$^{-1}$ (FWHM; 90\% confidence range).  A Gaussian model is a good fit to the data.

Figure~\ref{fig:Fe-Kalpha-cont} shows the confidence contours between the Fe-K$\alpha$ intensity (sum of two lines) and redshift.  The redshift of $0.01702^{+0.00059}_{-0.00060}$ is consistent, within 90\% errors, with the value ($z_*=0.017284$) measured by optical emission lines of stars in NGC~1275 \citep{Hitomi2017a}.  This shows that the center energies of the Hitomi detected lines are consistent with the neutral values of Fe-K$\alpha_1$ and Fe-K$\alpha_2$ lines at 6.404 and 6.391~keV, respectively, at the redshift of NGC~1275. However, the derived AGN redshift is inconsistent with that of the of the ambient ICM emission ($z_{\rm ICM}=0.01767\pm0.00003$; \cite{2016Natur.535..117H}).  Refer to \citet{Hitomi2017a} for more discussions about the redshift difference between NGC~1275 and ICM.

%\bibitem[Strauss et al.(1992)]{1992ApJS...83...29S} Strauss, M.~A., Huchra, J.~P., Davis, M., et al.\ 1992, \apjs, 83, 29 
%\bibitem[Hitomi Collaboration et al.(2016)]{2016Natur.535..117H} Hitomi Collaboration, Aharonian, F., Akamatsu, H., et al.\ 2016, \nat, 535, 117

In order to consider systematic errors which come from the SXS gain uncertainty and the spatial distribution of the ICM bulk motion, we next fit the 6.1--6.45~keV spectrum extracted from the event sets with neither the $z$ nor parabolic correction, with the same models and parameter settings as above. As shown in table~\ref{tab:Fe-Kalpha}, the Gaussian sigma becomes $5.2$--$14.4$~eV (90\% confidence range after including the instrumental broadening in the modeling) which corresponds to the 90\% confidence range of the velocity width of 600--1600~km~s$^{-1}$ (FWHM).  The line detection significance is still $\sim 5.4~\sigma$. The systematic increase of the line width is $\sim 1$~eV, and this is consistent with that due to the gain uncertainty of the SXS  and/or the region-by-region ICM bulk motion \citep{Hitomi2017a}. The redshift becomes $0.01733^{+0.00065}_{-0.00064}$ which is consistent, within statistical 90\% error, with that derived from the spectrum with the gain corrections, and with the optically-measured value ($z_*=0.017284$). Thus, even when the systematic errors are taken into account, the central energies of the Fe-K$\alpha$ lines are consistent with their neutral values. Hereafter, we thus utilize the velocity width range of $\sim 500$--1600~km~s$^{-1}$ (FWHM), taking into account the systematic uncertainties.

%--------------------------S3.1.2------------------------
\subsubsection{Spatial Distribution from SXS Data}\label{s3-1}
%--------------------------S3.1.2------------------------

The Perseus cluster possesses a system of atomic and molecular filaments that extend throughout the inner core of the cluster (see the detail in \S\ref{S4-1}).  The origin of this cold gas which is embedded within the hot ICM is still debated; it may be condensing via thermal instability from the hot ICM, or it may be cold gas from the galaxy that is uplifted into the ICM by AGN activity. Either way, as discussed more extensively below, irradiation of this cold gas by the hot ICM and the central AGN can produce spatially extended Fe-K$\alpha$ fluorescence.  We begin our exploration (and ultimate rejection) of this possibility for the observed Fe-K$\alpha$ fluorescence by using the imaging capability of the SXS to directly examine the spatial distribution of the Fe-K$\alpha$ emission. 

%%%%%%%%%%%%%%%%figure3%%%%%%%%%%%%%%%%%%%%%%%%
\begin{figure*}[t]
\begin{center}
\FigureFile(150mm,150mm){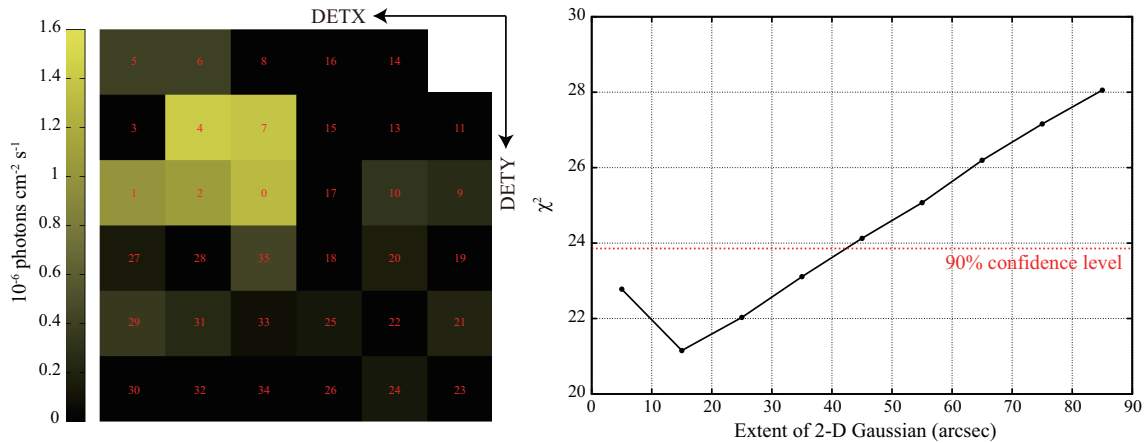}
\end{center}
\caption{Left: The SXS pixel distribution of the Fe-K$\alpha$ count rate. Right: A plot of extent of source versus $\chi^2$ values when the extent of the simulated 2-D Gaussian source, which is fitted to the Fe-K$\alpha$ map (left panel), is changed as 5, 15, ... , 65~arcsec. }
\label{fig:PixelDist}
\end{figure*}
%%%%%%%%%%%%%%%%figure3%%%%%%%%%%%%%%%%%%%%%%%%

We start by extracting spectra from individual pixels of the SXS microcalorimeter array. To select time intervals when the satellite pointing was stable, we use the GTI with a condition of the 1st Euler angle of 49.93--49.9325 degree and the 2nd angle of 48.4812--48.4826 degree. We use the same RMF and ARF as those utilized in \S3.1.1 (see \S2.1).  The individual pixel spectra are then fitted with the same model employed above, except that $\sigma_1$ and  $\sigma_2$ are fixed at 7.79~eV which is same as that in table~\ref{tab:Fe-Kalpha}.  As a result, we can derive a pixel distribution of count rate of the Fe-K$\alpha$ line photons as shown in figure~\ref{fig:PixelDist}(left). It includes $1\sigma$ uncertainties of $6\times10^{-7}$ and $(2-3)\times10^{-7}$~photons~cm$^{-2}$~s$^{-1}$ in the peak position (Pixel 4) and in the faint end (Pixel 9, 11, 19, 21, 23, 24, 26, 30, 32, and 34), respectively.We find that the Fe-K$\alpha$ count rate in the on-core pixel is highest as expected, and those in the surrounding 8 pixels are higher than off-AGN regions, consistent with the Hitomi PSF, as detailed below.

Using this observed distribution of the count rate of Fe-K$\alpha$ photons, we constrain the source extent by comparing with model distributions blurred by the 2-dimensional SXT Point Spread Function (PSF) as computed using the {\tt simx}v2.4.1 package.  We use the PSF 2-dimensional data file\footnote{\tt sxt-s-PSFimage\_150120.fits packed in the {\tt simx}v2.4.1 package} taken by the beam-line experiment of the flight SXT-S at ISAS/JAXA in 2014. Firstly, data from the Hitomi star-trackers are used to determine the precise location of NGC~1275 on the SXS array which we determine to be (DETX, DETY) = (4.85, 2.21).  Next, we simulate the SXS image of a point source of 6.4~keV line photons, producing a grid of models with different source locations.  We produce an 11$\times$11 grid of DETX and DETY with a step size of 0.2~pixels. 
Comparing these models to the observed Fe-K$\alpha$ count rate map using the $\chi^2$ statistic and making a confidence contour of DETX vs. DETY, we confirm that the best-fit location of the Fe-K$\alpha$ source becomes (DETX, DETY) =  (4.7$\pm$0.4, 2.5$\pm$0.4), consistent with the position of NGC~1275. 

Finally, we simulate a source that has an extended Gaussian profile, centered on NGC~1275, with $\sigma=5, 15, 25, ... , 65$~arcsec. The angular resolution of the SXS + SXT is not enough, and the  ratios between the Fe-K$\alpha$ and continuum count rates in the SXS pixel map are too low to limit complex source emission models which have effective extents smaller than the PSF. Hence, we choose a Gaussian as a simple model which can fit the data and constrain the source extent, considering a relatively smeared structure better than peakier models. The change in $\chi^2$ statistic as a function of extent is shown in figure~\ref{fig:PixelDist}(right).  Thus, the Hitomi SXS data show that the Fe-K$\alpha$ source is constrained to have an extent of less than $\sim42$~arcsec or $\sim 17$~kpc at the distance of NGC~1275.

%%%%%%%%%%%Table 3%%%%%%%%%%                                                                                                                                                   
\renewcommand{\arraystretch}{1}
\begin{table*}[t]
 \caption{Results of the spectral fits to the XMM-Newton, Chandra/HEG, and Chandra/ACIS spectra in the 4--8~keV band. }
 \label{all_tbl}
 \begin{center}
  \begin{tabular}{cccccc}
   \hline\hline
Component         & Parameter                & XMM-Newton/PN and MOS & XMM-Newton/MOS  & Chandra/HEG & Chandra/HEG  \\
  & & in 2001 & in 2006 & in 1999 & in 2000 \\
                  &                          &                   &             & \multicolumn{2}{c}{(Simultaneous fitting)}        \\ \hline
\texttt{constant} &$C_{\rm inst}$            & $0.85\pm0.07$               & ...                         & ...             & ...                            \\
\texttt{powerlaw} &$\Gamma$                  & $1.24\pm0.19$               & $1.50 \pm 0.14$             & $0.76\pm0.58$   & $2.01^{+0.48}_{-0.47}$         \\
                  &$N_{\rm PL}^{\rm a}$      & $1.35^{+0.50}_{-0.36}$      & $2.79^{+0.73}_{-0.58}$      & $0.56^{+0.90}_{-0.35}$ & $9.81^{+11.23}_{-5.20}$ \\
\texttt{zgauss}   &$E$~(keV)                 & \multicolumn{4}{c}{$6.40$ (fix)}     \\
                  &$\sigma$~(eV)             & \multicolumn{4}{c}{$10$ (fix)}       \\
                  & $z$                      & \multicolumn{4}{c}{$0.017284$ (fix)} \\
                  & $N_{\rm zgauss}^{\rm b}$ & $12.59^{+6.22}_{-5.90}$     &$13.26^{+6.01}_{-5.79}$      & \multicolumn{2}{c}{$2.35^{+6.74}_{-2.35}$} \\[1.5ex]
\hline $C$-statistics/d.o.f. &                     & 1418.03/1445                & 820.86/788                  & \multicolumn{2}{c}{615.14/789} \\
\hline\hline
Component       & Parameter                & \multicolumn{4}{c}{Chandra/ACIS in 2004 (Simultaneous fitting)} \\\hline 
%       &                & \multicolumn{4}{c}{ (Simultaneous fitting) } \\\hline
\texttt{apec}   & $T_{\rm e}$~(keV)              & \multicolumn{4}{c}{$4.22\pm0.09$}\\
                & $A$~(Solar)                      & \multicolumn{4}{c}{$0.47\pm 0.01$} \\
                & $z_{\rm apec}$           & \multicolumn{4}{c}{$0.017284$ (fix)} \\
                & $N_{\rm apec}^{\rm }$   & \multicolumn{4}{c}{$3.70\pm0.08$}\\[1.5ex] 
\texttt{zgauss} & $E$~(keV)                & \multicolumn{4}{c}{$6.40$ (fix)}\\
                & $\sigma$~(eV)            & \multicolumn{4}{c}{$10$ (fix)} \\
                & $z_{\rm zgauss}$         & \multicolumn{4}{c}{$=z_{\rm apec}$}\\
                & $N_{\rm zgauss}^{\rm b}$ & \multicolumn{4}{c}{$1.90^{+1.50}_{-1.49}$}\\[1.5ex]
\hline             $C$-statistics/d.o.f.         &  &  \multicolumn{3}{c}{3232.51/2999} \\
\hline\hline
  \end{tabular}
\end{center}
 {\small
   $^{\rm a}$ The \texttt{powerlaw} normalization at 1 keV, in units of 10$^{-3}$ photons~keV$^{-1}$~cm$^{-2}$~s$^{-1}$. \\
   $^{\rm b}$ The \texttt{zgauss} normalization, in units of $10^{-6}$~photons~cm$^{-2}$~s$^{-1}$. \\
   $^{\rm c}$ The \texttt{apec} normalization, in units of $(10^{-16}/4 \pi (D_{\rm A} (1+z))^{2}) \int n_{\rm e} n_{\rm H} dV$,
               where $D_{\rm A}$ is the angular size distance to the source (cm), $n_{\rm e}$ and $n_{\rm H}$ are the electron
               and H densities (cm$^{-3}$). \\
}
\label{tab:Fe-Kalpha_DiffSat}
\end{table*}
%%%%%%%%%%%Table 3%%%%%%%%%%%%  

%%%%%%%%%%%%%%%%figure4%%%%%%%%%%%%%%%%%%%%%%%%
\begin{figure}[t]
\begin{center}
\FigureFile(70mm,70mm){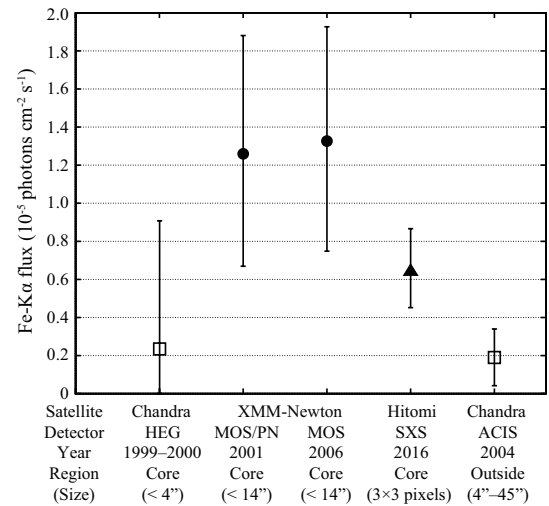}
\end{center}
\caption{Comparison of the Fe-K$\alpha$ fluxes among different satellites, instruments, years, and regions. Open squares, filled circles, and filled triangle show the Fe-K$\alpha$ fluxes obtained by Chandra, XMM-Newton, and Hitomi, respectively. Error bars show $90$\% confidence ranges. }
\label{fig:SeveralSat}
\end{figure}
%%%%%%%%%%%%%%%%figure4%%%%%%%%%%%%%%%%%%%%%%%%

%--------------------------S3.1.3------------------------
\subsubsection{Fe-K$\alpha$ Comparison with Chandra and XMM-Newton}\label{s3-1-2}
%--------------------------S3.1.3------------------------

Historical variability of the Fe-K$\alpha$ line intensity is useful to diagnose the location of the fluorescing matter. We begin with a comparison to the archival XMM-Newton data taken in 2001 and 2006. The procedures of the data reduction as described below are based on the XMM-Newton ABC guide\footnote{https://heasarc.gsfc.nasa.gov/docs/xmm/abc/}. The raw PN and MOS data are reprocessed using pipelines, \texttt{epchain} and \texttt{emchain}, respectively. 

We filter the duration where the background activity is high, judged from the background light curve in the 10--12~keV band. We also impose the event pattern selection, PATTERN $\leq$ 4 (single and double events) for PN, and PATTERN $\leq$ 12 (single, double, triple, and quadruple events) for MOS. To determine the source region, pile-up is checked with the SAS tool \texttt{epatplot}, and we find it is significant in the vicinity of the AGN. Hence, depending on its strength, on-core $7''$--$14''$ and $8''$--$14''$-annular regions are adopted for the 2001 and 2006 data, respectively, to exclude the pixels suffering from pile-up. Note that the PN data in 2006 is not used because the pile-up effects the entire extraction region for NGC~1275. The background spectra including the ICM emission is estimated from a $60$--$63''$ annular region following \citet{2013PASJ...65...30Y}. We combine the spectra from the two individual MOS detectors. 

%%%%%%%%%%%%%%%%figure5%%%%%%%%%%%%%%%%%%%%%%%%
\begin{figure*}[t]
\begin{center}
\FigureFile(150mm,150mm){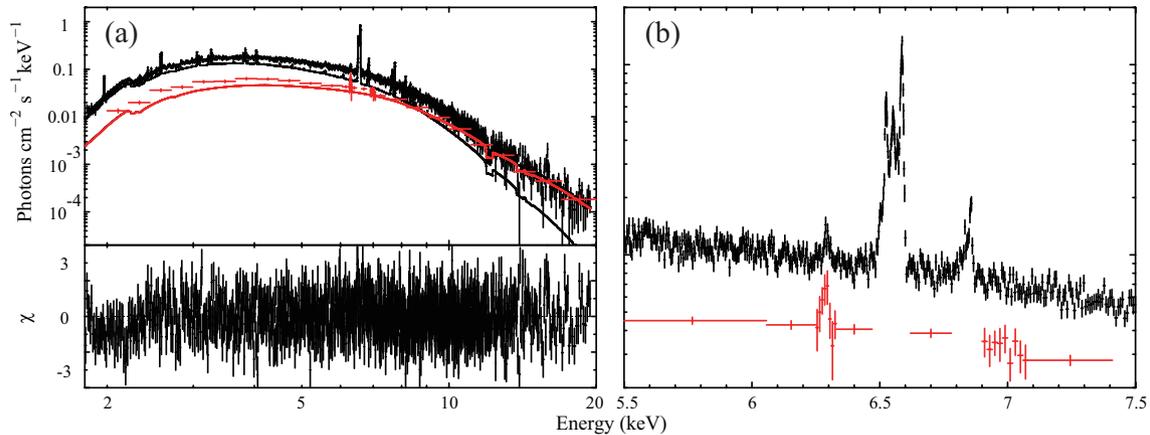}
\end{center}
\caption{(a) The broad-band spectrum obtained by the SXS on-core 9 pixels, fitted with the model of the ICM plus AGN emission. Black is the on-core 9-pixels SXS spectrum, while red is the spectrum of NGC~1275 derived by the PSF photometry method. The black and red solid lines are the best-fit \texttt{bbvapec} and \texttt{pegpwrlw} component, respectively, derived by the fit to only the SXS on-core 9-pixels spectrum.  (b) Enlargement of panel (a) around the Fe-K$\alpha$ line region without the models. }
\label{fig:SXSbroadband}
\end{figure*}
%%%%%%%%%%%%%%%%figure5%%%%%%%%%%%%%%%%%%%%%%%%

Our main focus is accurate measurement of the Fe-K$\alpha$ flux, and thus we have to determine the continuum level with the least bias. For that purpose, we analyze the 4--8~keV spectra to reduce the systematic uncertainty caused by the ICM emission subtraction. The spatial variation of the ICM emission also makes suitable background subtraction difficult. The RMFs and ARFs are generated in a standard manner appropriate to a point source. We fit these data with a model consisting of a powerlaw continuum and a Gaussian Fe-K$\alpha$ emission line (\texttt{powerlaw +  zgauss}). When fitting the PN and MOS spectra in 2001, the systematic uncertainty between the two is taken into consideration by applying the \texttt{constant} model to the total ($C_{\rm inst}$ in table~\ref{tab:Fe-Kalpha_DiffSat}). The photon index and normalization of the \texttt{powerlaw}, and the normalization of the\texttt{zgauss} are left free, while the line energy, its width, and the redshift of \texttt{zgauss} are fixed at 6.4~keV, 10~eV, and 0.017284, respectively. Table~\ref{tab:Fe-Kalpha_DiffSat} summarizes the fitting results. The Fe-K$\alpha$ fluxes are measured to be $(12.59^{+6.22}_{-5.90}) \times 10^{-6}$ ~photons~cm$^{-2}$~s$^{-1}$ in 2001, and $(13.26^{+6.01}_{-5.79}) \times 10^{-6}$~photons~cm$^{-2}$~s$^{-1}$ in 2006 (errors are 90\% confidence). The line fluxes are in good agreement with that in 2016 obtained from the Hitomi data despite the significant brightening of the AGN continuum (\texttt{powerlaw}) emission indicating a lack of immediate response of the line to continuum changes. 

%\bibitem[Yamazaki et al.(2013)]{2013PASJ...65...30Y} Yamazaki, S., Fukazawa, Y., Sasada, M., et al.\ 2013, \pasj, 65, 30 
%https://heasarc.gsfc.nasa.gov/docs/xmm/sas/USG/epicpileup.html

To extract the AGN spectra from the HEG data, we run \texttt{chandra\_repro} with the default parameter set. There is no evidence for pile-up using the CIAO tool \texttt{pileup\_map}. The source region is a rectangle surrounding dispersed X-rays with the width of $\sim 4.8''$, whereas that for the background including the ICM emission is two rectangles with the width of $\sim 22''$ located on either side along the source region. The RMFs and ARFs are produced by assuming a point source. To get a stronger constraint on the Fe-K$\alpha$ flux, we simultaneously fit the 4--8~keV spectra in 1999 and 2000 using the model of \texttt{powerlaw + zgauss}. We determine the continuum component (\texttt{powerlaw}) independently in the two spectra but tie the Fe-K$\alpha$ line intensity. The resultant intensity is $(2.35^{+6.74}_{-2.35}) \times 10^{-6}$~photons~cm$^{-2}$~s$^{-1}$, consistent with those in the XMM-Newton (2001 and 2006) and Hitomi (2016) observations.

In the same manner as for the Chandra/HEG data, we reprocess the Chandra/ACIS data, and find that while the nuclear region ($< 4''$) is badly affected by  pile-up, the outer regions are not affected. Hence, the Chandra/ACIS allows us to investigate if there is an Fe-K$\alpha$ source other than the AGN. The SXS pixel distribution of the Fe-K$\alpha$ flux has already limited the extent of the Fe-K$\alpha$ emitter to be $\lesssim 42''$, so we use the Chandra/ACIS data to constrain the flux in the $4''$--$45''$ annular region around the AGN. The background region is a $50''$ circle, where no other source is present. The 4--8~keV spectra are extracted from all the Chandra/ACIS archival data listed in table~\ref{tab:OtherSat}. With RMFs and ARFs optimized for the extended source, we perform a simultaneous fitting to the 11 spectra with a model consisting of thermal plasma emission from the ICM and a neutral Fe-K$\alpha$ line (\texttt{apec + zgauss}).  All parameters in the ICM model except for the redshift of $0.017284$ (i.e., the electron temperature, elemental abundance, and normalization) are left free, while \texttt{zgauss} is treated in the same manner in the HEG analysis. The neutral Fe-K$\alpha$ line flux is constrained to be $(1.90^{+1.50}_{-1.49})\times10^{-6}$~photons~cm$^{-2}$~s$^{-1}$. Although the detection is significant at  $\sim 97$\% confidence level,  the resultant flux is significantly smaller than those measured with XMM-Newton/MOS and PN, Chandra/HEG, and Hitomi/SXS. We conclude that the Fe-K$\alpha$ emission in the region in the  $4''-42"$ annulus from the core does not significantly contribute to the Hitomi detection and thus the Fe-K$\alpha$ flux observed with the SXS comes mainly from the nucleus region within a $4''$ radius, or $\sim 1.6$~kpc from NGC~1275. Hence, the Fe-K$\alpha$ emitter is likely located within the circumnuclear material.

%--------------------------S3.2------------------------
\subsection{Broad-Band Spectral Analysis}\label{s3-2}
%--------------------------S3.2------------------------

%--------------------------S3.2.1------------------------
\subsubsection{Direct fit to the SXS on-core 9-pixels spectrum}\label{s3-2-1}
%--------------------------S3.2.1------------------------

%%%%%%%%%%%%%%%%table4%%%%%%%%%%%%%%%%%%%%%%%%
\begin{table*}[t]
\caption{Results of the fits to the SXS on-core 9-pixels spectrum and the PSF-photometry spectrum in the baseline case, and in the cases utilizing the ARFs correspoding to the ground and orbital calibration.}
\begin{center}
\label{tab:SXSbroadband}
\begin{tabular}{ccccc}
\hline\hline
   Component & Parameter  & Baseline & Ground calibration & Orbital calibration \\\hline 
\multicolumn{5}{c}{Direct fit to the SXS on-core 9-pixels spectrum} \\[1.5ex]
  \texttt{pegpwrlw} & $\Gamma$
  			& $1.60^{+0.13}_{-0.15}$ & $1.99^{+0.07}_{-0.10}$ & $1.71^{+0.10}_{-0.11}$\\
			
			&$F_{\rm 2-10~keV}^{\rm a}$
			&$2.31^{+0.40}_{-0.35}$ & $3.93^{+0.42}_{-0.49}$ & $3.05^{+0.41}_{-0.39}$ \\[1.5ex]
			
  \texttt{zgauss}$\times2$ & EW~(eV) & $26.5^{+3.7}_{-3.1}$ & $16.4^{+1.6}_{-1.1}$ & $20.7^{+2.3}_{-1.9}$ \\[1.5ex] 
			  
 \hline  $C$-statistics/d.o.f. &  &  10562.2/11120 & 10559.6/11120 & 10560.9/11120\\\hline\hline
 
\multicolumn{5}{c}{Fit to the PSF-photometry spectrum} \\[1.5ex]
  \texttt{pegpwrlw} & $\Gamma$
  			& $1.83 \pm 0.04$ & $1.79 \pm 0.04$ & $1.76 \pm 0.04$\\
			
			&$F_{\rm 2-10~keV}^{\rm a}$
			&$2.97 \pm 0.06$ & $2.85 \pm 0.06$ & $2.98 \pm 0.06$ \\[1.5ex]
			
  \texttt{zgauss}$\times2$ & EW~(eV) & $20.3 \pm 0.3$ & $20.9 \pm 0.3$ & $19.8 \pm 0.3 $ \\[1.5ex] 
			
\hline $\chi^2$/d.o.f. &  &  21.0/39 & 25.2/39 & 20.1/39\\

\hline
\end{tabular}
\end{center}
$^{\rm a}$: The 2--10 keV flux of the \texttt{pegpwrlw} in units of $10^{-11}$ erg cm$^{-2}$ s$^{-1}$.
\end{table*}
%%%%%%%%%%%%%%%%table4%%%%%%%%%%%%%%%%%%%%%%%%

We now turn to a broad-band spectral analyses of the SXS data in order to constrain the AGN continuum. This requires a careful decomposition of the spectrum into ICM and AGN components; indeed, the spectrum of the AGN is important for any detailed studies of the ICM emission. Because the spectral continuum of the AGN is different from that of the ICM, it is effective to study the broadest band as possible. In spite of its being, nominally, a soft X-ray instrument, the combination of the SXT+SXS possesses sufficient sensitivity to detect NGC~1275 at $\sim20$~keV. At such high energies the AGN dominates the SXS signal (see figure~\ref{fig:SXSbroadband}).  The fact that the SXS gate-valve was still closed during this observation limits the analysis to energies above 1.8~keV.  We use the same RMF as in \S3.1.1 (see \S2.1). We use the diffuse and point-like source ARFs for the ICM and AGN model component, respectively. The ARF for the AGN model is same as that used in \S3.1.1 (see \S2.1).

We perform the direct fit to the 1.8--20~keV spectrum from the $3\times3$ pixels (\#0--8) region of the SXS centered on the AGN, which is same as that in \S3.1.1, with a model consisting of ICM and AGN components. Both the ICM and AGN components are modified by the effects of Galactic absorption (described using the XSPEC model {\tt TBabs}) with a column density of $1.38\times10^{21}$~cm$^{-2}$ derived from the Leiden/Argentine/Bonn (LAB) Survey of Galactic HI \citep{Kalb05}. The ICM emission is modeled with {\tt zashift*bvvapec}, which is the redshifted thermal plasma model including thermal and turbulent line broadening, together with the {\tt AtomDB} version 3.0.9
\footnote{Because treatments of some Fe-K lines are updated from {\tt AtomDB} version 3.0.8, finally we use version 3.0.9 in the present paper. The APEC version 3.0.8 and 3.0.9 produce just small difference in the photon index of $\sim0.05$ in the AGN spectral fitting.}. 
Avoiding the mismatch of line ratio due to resonance scatterings \citep{Hitomi2017c}, we exclude Fe~XXV~He$\alpha$ resonance line from the model, and instead we introduce one Gaussian model to represent this resonance line. Hence, the model fitted to the ICM is {\tt TBabs*zashift*bvvapec + gauss}. 
Light elements, He through to Al, are assumed to have solar abundances (as defined by \cite{2009M&PSA..72.5154L}) because the relevant lines of these elements are out of the SXS energy range. The abundance of other elements are allowed to be non-solar, but those of P, Cl, K, Sc, Ti, V, Co, Cu, and Zn are all tied to that of Fe, following \citet{Hitomi2017b}. The AGN component is modeled by a powerlaw plus the iron fluorescence lines (Fe-K$\alpha_1$+Fe-K$\alpha_2$), {\tt pegpwrlw + zgauss + zgauss}, where the normalization ratio of two Gaussians is fixed to $1:2$ as described in \S3.1.1.  We perform the spectral fit in the energy range of 1.8--20~keV with $C$-statistics. The spectrum is grouped so as to have at least one photon in each channel. 

%%%%%%%%%%%%%%%%table5%%%%%%%%%%%%%%%%%%%%%%%%
\begin{table}[t]
\caption{The abundance values of the ICM modeled by {\tt bvvapec} in the direct fit to the SXS on-core 9-pixels spectrum$^{\rm a}$.}
\begin{center}
\label{tab:ICM}
\begin{tabular}{cc}
\hline\hline
   Element & Abundance (Solar)  \\\hline 

Si & $0.67^{+0.13}_{-0.12}$\\ 
P & $ = A_{\rm Fe}$\\ 
S & $0.75^{+0.09}_{-0.07}$\\ 
Cl & $ = A_{\rm Fe}$\\ 
Ar & $0.64^{+0.08}_{-0.09}$\\ 
K & $ = A_{\rm Fe}$\\ 
Ca & $0.71^{+0.10}_{-0.08}$\\
Sc & $ = A_{\rm Fe}$\\ 
Ti & $ = A_{\rm Fe}$\\ 
V & $ = A_{\rm Fe}$\\ 
Cr & $0.35^{+0.23}_{-0.22}$\\
Mn & $0.54^{+0.35}_{-0.33}$\\
Fe & $0.65^{+0.06}_{-0.05}$\\
Ni & $0.71^{+0.14}_{-0.13}$\\
Cu & $ = A_{\rm Fe}$\\
Zn & $ = A_{\rm Fe}$\\ 		
			
\hline \hline
\end{tabular}
\end{center}
$^{\rm a}$ $A_{\rm Fe}$ is the Fe abundance in a unit of the Solar abundance.
\end{table}
%%%%%%%%%%%%%%%%table5%%%%%%%%%%%%%%%%%%%%%%%%

As a result, the SXS spectrum is well fitted by this model with $C$-statistics/d.o.f.$=10562.2/11120$, as shown in figure~\ref{fig:SXSbroadband}(a). The direct fit results of the AGN parameters are summarized in the Baseline column of table~\ref{tab:SXSbroadband}. The ICM parameters become $z_{\rm ICM}= 0.01729^{+0.00002}_{-0.00004}$, $T_{\rm e}=3.82 \pm 0.07$~keV, and $V_{\rm turb}=158.6^{+11.7}_{-11.9}$~km~s$^{-1}$ which are almost consistent with those reported by Hitomi Collaboration (2017a, b). The ICM abundance values of the various elements are summarized in table~\ref{tab:ICM}, and their ratios are confirmed to be consistent with those reported in \citet{Hitomi2017d}. Hence, we do not furthermore discuss the ICM parameters in the present paper.

%\bibitem[Kalberla et al.(2005)]{Kalb05} Kalberla, P.~M.~W., Burton, W.~B., Hartmann, D., et al.\ 2005, \aap, 440, 775 
%\bibitem[Lodders et al.(2009)]{Lodd09} Lodders, K., Palme, H., \& Gail, H.-P.\ 2009, Landolt B{\"o}rnstein,  

%--------------------------S3.2.2------------------------
\subsubsection{Analyses of the PSF-photometry spectrum}\label{s3-2-2}
%--------------------------S3.2.2------------------------

Although the statistical errors on the AGN spectral slope and flux are rather small as shown in  Table~\ref{tab:SXSbroadband}, they are correlated with the ICM parameters, and potentially, susceptible to systematics resulting from uncertainties in the calibration and modeling. In fact, the powerlaw photon index is distributed widely from 1.4 to 2.2 when we use the different gain calibration, RMF size, and ARF effective area 
\footnote{Note that, as described in \S2.1, we use the additional gain calibration in addition to that used in the default pipeline processing of the SXS data. Furthermore, there are four options on the RMF size in the SXS RMF generator {\tt sxsmkrmf}. These RMFs are different in not only size but also some response issues considered in the calculation. See \S3.2.2 for the systematic errors due to the ARF effective area.}. 
This is because the AGN component is buried in the ICM emission especially below 6~keV and hence, the spectral slope can be significantly affected by a small change of the calibration and modeling. Therefore, we perform a model-independent evaluation of the AGN spectrum via the image analysis, employing the PSF photometry on the SXS pixel maps derived in multiple narrow energy bands (see Appendix~1 for further details of the creation of the PSF-photometry spectrum). The derived PSF-photometry spectrum is shown in figure~\ref{fig:SXSbroadband}(a). It is successfully determined up to 20~keV, and the Fe-K$\alpha$ line also clearly appears as shown in figure~\ref{fig:SXSbroadband}(b). 

%%%%%%%%%%%Table 5%%%%%%%%%%                                                                                                                                                   
\renewcommand{\arraystretch}{1}
\begin{table*}[t]
 \caption{Results of the fits to the 0.5--7~keV Chandra/HEG and 0.7--8~keV XMM-Newton spectra.}
 \label{all_tbl}
 \begin{center}
  \begin{tabular}{cccccc}
   \hline\hline
Component         & Parameter                &  Chandra/HEG & Chandra/HEG  & XMM-Newton/PN and MOS  & XMM-Newton/MOS  \\
  & & in 1999 & in 2000 & in 2001 & in 2006 \\
\hline
\texttt{pegpwrlw} &$\Gamma$                  & $1.66\pm0.07$               & $2.12 \pm 0.05$             & $1.74\pm0.02$   & $1.75^{+0.02}_{-0.01}$                   \\
                  &$F_{\rm 2-10~keV}^{\rm a}$      & $1.05 \pm 0.07$      & $2.44 \pm 0.11$      & $1.17\pm0.03$  & $1.56\pm0.03$ \\[1.5ex]
\hline $C$-statistics/d.o.f. &                     & 2857.4/3537              & 2937.1/3655                  & 2707.7/2768 &  1591.5/1447\\
\hline\hline
  \end{tabular}
\end{center}
$^{\rm a}$: The 2--10 keV flux of the \texttt{pegpwrlw} in units of $10^{-11}$ erg cm$^{-2}$ s$^{-1}$.
\label{tab:broadband_DiffSat}
\end{table*}
%%%%%%%%%%%Table 5%%%%%%%%%%%% 

We fit the PSF-photometry spectrum with an absorbed powerlaw model with two Gaussians modeled as {\tt TBabs*(pegpwrlw + zgauss + zgauss)}
\footnote{Because we do not use {\tt AtomDB} in the fit to the PSF-photometry spectrum, the results in this subsection are not related to the {\tt AtomDB} version update.}. 
We fix the Galactic absorption column density to that used in the direct fit in \S3.2.1. Because the binsize of the PSF-photometry spectrum is much larger than that of the SXS on-core 9-pixels spectrum, we fix the parameters of two Gaussians at the best-fit values with the gain corrections in table~\ref{tab:Fe-Kalpha_DiffSat}. Because the errors, given by the PSF photometry method (see Appendix~1), do not follow a Poisson distribution, we use $\chi^2$ statistics in place of $C$-statistics. As shown by the Baseline column in table~\ref{tab:SXSbroadband}, the fit is successful with $\chi^2$/d.o.f$=21.0/39$, and the \texttt{pegpwrlw} photon index and flux in 2--10~keV become $1.83\pm0.04$ and $(2.97 \pm 0.06) \times10^{-11}$~erg~cm$^{-2}$~s$^{-1}$, respectively. The EW of the Fe-K$\alpha$ line against the AGN continuum becomes $20.3 \pm 0.3$~eV. Incorporating the PSF photometry method, we thus obtain the AGN spectrum which is slightly steeper than that derived by the direct fit in \S3.2.1. Note that these AGN parameters based on the PSF photometry manner are employed in the other Hitomi papers of the Perseus cluster of galaxies (Hitomi Collaboration 2017a, b, c, d, e).

%--------------------------S3.2.3------------------------
\subsubsection{Broad-band comparison with Chandra and XMM-Newton including systematic errors}\label{s3-2-3}
%--------------------------S3.2.3------------------------

Finally, in order to compare the AGN parameters in 2016 with those in the previous occasions by the different satellites, we check parameter changes due to the effective area uncertainties, which mainly produce large systematic errors. The effective area calculated by the \texttt{aharfgen} is different from that measured in the ground calibration by up to $\sim7$\% (Hitomi Collaboration 2017a, e), and that investigated in the orbital calibration by utilizing the Crab ratio by up to $\sim10$\% (\cite{Tsujimoto17}; Hitomi Collaboration 2017a, e). Then, we modify the ARF, according to the differences from the ground and orbital calibration individually, and again perform the fits to the SXS on-core 9-pixels spectrum and the PSF-photometry spectrum. Results are summarized in table~\ref{tab:SXSbroadband}. Consequently, considering the systematic uncertainties of the direct fits, which are wider than those in the fits to the PSF-photometry spectrum, the parameter range of the powerlaw photon index is determined to be 1.45--2.06. The 2--10 keV flux range is $(1.96$--$4.35)\times10^{-11}$~erg~cm$^{-2}$~s$^{-1}$, which is comparable to the AGN brightness around 1980--90s \citep{2015MNRAS.451.3061F}. The range of the Fe-K$\alpha$ EW against the AGN continuum becomes 15.3--30.2~eV.  

How different are the AGN spectral shape and flux obtained by the SXS in 2016 from those in the past X-ray observations? First, we check the Chandra/HEG 0.5--7~keV spectra obtained in 1999 and 2000 (see table~1). The Chandra/HEG source and background spectra, the RMFs and ARFs are all same as those in \S3.1.3, but the energy band is changed to 0.5--7~keV. We fit the HEG spectra with the model of {\tt TBabs*(pegpwrlw + zgauss)}, where the parameter settings of \texttt{TBabs} and \texttt{pegpwrlw} are same as in \S3.2.2, while those of \texttt{zgauss} are fixed at the values with the gain corrections in table~\ref{tab:Fe-Kalpha_DiffSat} except for its normalization left free. As shown in table~\ref{tab:broadband_DiffSat}, the fits are successful, and the 2--10~keV flux in 1999 is significantly lower than 2016, while that in 2000 is compatible to 2016. 

The XMM-Newton broad-band spectrum in 2001 and 2006 were already reported by \citet{2003ApJ...590..225C} and \citet{2013PASJ...65...30Y}, respectively. However, as shown in \S3.1.3, we utilize the different source- and background-extraction regions to avoid pile-up effects, and hence, we reanalyze the 0.7--8~keV spectra in both 2001 and 2006. The source and background spectral files, the RMS and ARFs are identical to those in \S3.1.3, and the energy band is broadened to 0.7--8~keV. We fit the MOS and PN spectra simultaneously in 2001, while only the MOS spectrum in 2006, using the model of {\tt TBabs*(pegpwrlw + zgauss)} with the same parameter settings as those in the Chandra/HEG 0.5--7 keV analyses. The fits are both successful as shown in table~\ref{tab:broadband_DiffSat}, and their 2--10~keV fluxes become between those by Chandra/HEG in 1999 and 2000, and significantly lower than that by SXS in 2016. Comparing the results by the Chandra/HEG, XMM-Newton and the SXS with the systematic errors, the relation between the photon index and the flux of the AGN continuum is confirmed to be consistent with the trend of ``steeper when brighter", which is often seen in the AGN slope variability. 

%==============S4===============
\section{Discussion}\label{s4}
%==============S4===============

%--------------------------------------------------
\subsection{The origin of the Fe-K$\alpha$ line from NGC~1275}\label{S4-1}
%--------------------------------------------------

The Fe-K$\alpha$ velocity width of $\sim500$--1600~km~s$^{-1}$ (\S3.1.1) is significantly below that of the broad H$\alpha$ emission line ($\sim 2750$~km~s$^{-1}$) reported by \citet{1997ApJS..112..391H}. Applying simple Keplerian arguments (assuming that the black hole mass of $8\times 10^8~M_\odot$ dominates the potential and that the distribution is an edge-on disk) suggests that the fluorescing matter is $\sim1.4$--$14$~pc from the black hole --- the inclusion of the stellar mass and finite inclination effects would further increase this distance estimate.  We conclude that the matter responsible for the Fe-K$\alpha$ emission line in NGC~1275 is exterior to the BLR, possibly being associated with the dusty torus of unification schemes, or the circumnuclear molecular disk \citep{2013MNRAS.429.2315S}. 
Furthermore, the inner parts of the cluster-scale CO clouds illuminated by the ICM X-ray emission should be also considered as a possible site, because they were reported to have a velocity field which matches that derived in the present study \citep{2006A&A...454..437S}. The cluster-scale origin of Fe-K$\alpha$ line emission was originally predicted by \citet{Chur98}, and discussed as well by \citet{2015MNRAS.451.3061F}.

%\bibitem[Churazov et al.(1998)]{1998MNRAS.297.1274C} Churazov, E., Sunyaev, R., Gilfanov, M., Forman, W., \& Jones, C.\ 1998, \mnras, 297, 1274 

The constraint imposed by the SXS pixel map (\S3.1.2) and the Chandra imaging (\S3.1.3), namely that the line emission is within 1.6~kpc of NGC~1275, firmly places the fluorescing material in the region where the AGN continuum (as opposed to the ICM emission) is the dominant source of irradiation that drives the fluorescence. Thus, we can essentially rule out dominant contributions from the extended CO clouds illuminated by the ICM X-ray emission. 
Although the CO molecular clouds in the centrally concentrated region within 1.6~kpc \citep{2006A&A...454..437S} could be illuminated by the AGN emission, we confirm that Fe-K$\alpha$ flux from them is too small (the EW is just less than 0.15~eV) to explain the Hitomi observed Fe-K$\alpha$ flux, by incorporating the Monte-Carlo simulation (see Appendix~2 for details). As Chandra might have detected the Fe-K$\alpha$ line from the 4--45 arcsec region with a $\sim 97$\% confidence level, a relatively small fraction of the line flux detected by Hitomi could be produced by the CO gas irradiated by collimated and beamed X-rays from a relativistic jet observed in NGC~1275 \citep{Abdo09}. This possibility is discussed in \S4.2.

%\bibitem[Ho et al.(1997)]{1997ApJS..112..391H} Ho, L.~C., Filippenko, A.~V., Sargent, W.~L.~W., \& Peng, C.~Y.\ 1997, \apjs, 112, 391
%\bibitem[Scharw{\"a}chter et al.(2013)]{2013MNRAS.429.2315S} Scharw{\"a}chter, J., McGregor, P.~J., Dopita, M.~A., \& Beck, T.~L.\ 2013, \mnras, 429, 2315
%\bibitem[Salom{\'e} et al.(2006)]{2006A&A...454..437S} Salom{\'e}, P., Combes, F., Edge, A.~C., et al.\ 2006, \aap, 454, 437 

The Fe-K$\alpha$ line flux is found to be almost constant over a 16 year duration from the 1999--2000 Chandra to the 2016 Hitomi datapoint (\S3.1.3), despite a flux change of the powerlaw continuum by an order of magnitude shown in \citet{2015MNRAS.451.3061F}. The most obvious interpretation for the lack of line variability is to invoke light travel time effects in an extended source. Depending upon the precise geometry, the lack of time variability and the width of the Fe-K$\alpha$ lines are consistent with source of at least  $\sim5$~pc. Of the known structures within the NGC~1275 system, we conclude that the most likely origin of the fluorescent Fe-K$\alpha$ line is the putative dusty torus, or the observed circumnuclear molecular disk which extends to distances of 50--100~pc from the black hole \citep{2013MNRAS.429.2315S}. However, the dusty torus, if it is present, would have a relatively low-column density and/or low-covering fraction, and be rather different from that assumed in other works on the Fe-K$\alpha$ line in Seyfert galaxies, because the EW of the Fe-K$\alpha$ line in NGC~1275 is $\gtrsim5$ times less than what typical Seyferts have. This is consistent with the picture that AGNs with low Eddington ratios (which may include NGC~1275 with $L_B/L_{\rm Edd} \sim 3\times 10^{-4}$; \cite{2007ApJ...658..815S}) have low covering fractions of the dusty torus (e.g., \cite{2016ApJ...831...37K}; \cite{2017NatAs...1..679R}; \cite{2017Natur.549..488R}). This point is also discussed in \S4.3. 

%\bibitem[Ramos Almeida \& Ricci(2017)]{2017NatAs...1..679R} Ramos Almeida, C., \& Ricci, C.\ 2017, Nature Astronomy, 1, 679 

If we assume that the structure is Compton-thin, we can ``count'' fluorescing atoms to obtain an estimate of the mass of cold gas in this structure.  Following the simple analytic arguments of \citet{2000ApJ...540..143R} to relate the (average) EW of the fluorescent line $W_{\rm Fe}$ to the column density $N_H$, iron abundance $Z_{\rm Fe}$ and covering fraction $f_{\rm cov}$ of the fluorescing material as seen from the AGN, we obtain
\begin{equation}
W_{\rm Fe}\approx 65~f_{\rm cov}\left(\frac{Z_{\rm Fe}}{Z_\odot}\right)\left(\frac{N_H}{10^{23}~{\rm cm}^{-2}}\right)~{\rm eV}
\end{equation}
Taking the Hitomi value of $W_{\rm Fe}\approx 20$~eV, which is derived against only the AGN continuum (table 4), and assuming solar abundances for the circumnuclear matter, we conclude that $N_Hf_{\rm cov}\approx 3.0\times 10^{22}$~cm$^{-2}$.  Placing this matter at a distance of $r=100$~pc from the AGN, the resulting total mass of gas is $M=4\pi r^2f_{\rm cov}N_H\mu m_p$, which evaluates to $M\sim 4\times 10^7~{\rm M}_\odot$.  To put this number into perspective, if this material were to undergo efficient accretion into the central SMBH, it could power the observed AGN for about a Hubble time.  If the accretion were highly inefficient (say 0.1\%), this quantity of gas could power the observed AGN for approximately an ICM core cooling time.  We note that \citet{2013MNRAS.429.2315S} estimated the electron density of the [Fe II] emitters to be $\sim 4000$ cm$^{-3}$.  For this to be consistent with our conclusions requires either a small covering fraction ($f_{\rm cov}\approx 0.02$), a clumpy or shell-like medium, or both.  

%\bibitem[Ikeda et al.(2009)]{2009ApJ...692..608I} Ikeda, S., Awaki, H., \& Terashima, Y.\ 2009, \apj, 692, 608
%\bibitem[Reynolds et al.(2000)]{2000ApJ...540..143R} Reynolds, C.~S., Nowak, M.~A., \& Maloney, P.~R.\ 2000, \apj, 540, 143 
%\bibitem[Scharw{\"a}chter et al.(2013)]{2013MNRAS.429.2315S} Scharw{\"a}chter, J., McGregor, P.~J., Dopita, M.~A., \& Beck, T.~L.\ 2013, \mnras, 429, 2315

%--------------------------------------------------
\subsection{Consideration of Multi-Wavelength Results}\label{S4-2}
%--------------------------------------------------

NGC~1275 has a relatively high 12~$\mu$m luminosity from the central $\sim$100~pc region  \citep{2016ApJ...822..109A}, and the ratio between the X-ray and mid-infrared (MIR) is consistent with what expected from the X-ray and MIR luminosity relation in AGN with normal dusty tori \citep{2017ApJ...835...74I}. Hence, the MIR luminosity apparently supports the presence of a normal dusty torus, and that picture looks inconsistent with the picture obtained in \S\ref{S4-1}. However, because NGC~1275 is a well-known GeV/TeV gamma-ray emitter (\cite{Abdo09, Alek12}), the jet emission could strongly contribute to the MIR band, and a large fraction of the MIR luminosity could be explained by the jet emission. If so, the normal dusty torus is not required, and the picture is thus consistent with that in \S\ref{S4-1}. To examine whether or not the jet emission significantly contributes to the MIR band, we need further studies of the spectral energy distribution (SED) of the jet emission. 

If a strongly beamed jet X-ray emission, which is out of our line of sight, could illuminate a part of CO molecular clouds extended outside the AGN \citep{2006A&A...454..437S}, a high Fe-K$\alpha$ line flux might be generated, even though the solid angle of the clouds from the central engine is small. From the Gamma-ray observations, the jet power of NGC~1275 has been estimated to be $\sim 10^{44}$~erg~s$^{-1}$ \citep{Abdo09}, or higher \citep{Tave14}, and hence, the X-ray luminosity of the beamed jet reaches $\gtrsim 10^{43-44}$~erg~s$^{-1}$. This could produce the Fe-K$\alpha$ line flux enough to explain EW$\sim20$~eV against the jet X-ray continuum with the luminosity of $\sim 10^{43}$~erg~s$^{-1}$. Thus, this picture may explain all or some fraction of the observed Fe-K$\alpha$ line flux with neither a dusty torus nor a circumnuclear molecular disk in NGC~1275. In order to test this, understandings of the SED of the jet component are important as well as detailed studies of the velocity field of the extended molecular clouds.

%\bibitem[Asmus et al.(2016)]{2016ApJ...822..109A} Asmus, D., H{\"o}nig, S.~F., \& Gandhi, P.\ 2016, \apj, 822, 109 
%\bibitem[Ichikawa et al.(2017)]{2017ApJ...835...74I} Ichikawa, K., Ricci, C., Ueda, Y., et al.\ 2017, \apj, 835, 74 
%\bibitem[Abdo et al.(2009)]{Abdo09} Abdo, A.~A., Ackermann, M., Ajello, M., et al.\ 2009, \apj, 699, 31
%\bibitem[Aleksi{\'c} et al.(2012)]{Alek12} Aleksi{\'c}, J., Alvarez, E.~A., Antonelli, L.~A., et al.\ 2012, \aap, 539, L2 
%\bibitem[Tavecchio \& Ghisellini(2014)]{Tave14} Tavecchio, F., \& Ghisellini, G.\ 2014, \mnras, 443, 1224 

Furthermore, we consider a possible contribution of proton-induced Fe-K$\alpha$ line. Such an origin has been suggested to explain the diffuse 6.4 keV line at the Galactic center \citep{2015ApJ...807L..10N}. At the central region of NGC~1275, cosmic-ray protons could be enhanced by jet-induced shock acceleration \citep{2016AN....337...47K}. By assuming a proton energy spectrum with an index of $-2.5$ and an energy density of 20~eV cm$^{-3}$ in 0.1--1000~MeV, \citet{2015ApJ...807L..10N} estimated the intensity of proton-induced Fe-K$\alpha$ line to be $I_{\rm 6.4 keV}=6.6\times10^{35}\left({N_{\rm H}}/{10^{24}~{\rm cm}^{-2}}\right)\left({U_p}/{20~{\rm eV~cm}^{-3}}\right)\left({S}/{1~{\rm pc}^2}\right)$~erg~s$^{-1}$. Because the Fe-K$\alpha$ flux derived by the SXS corresponds to the much higher luminosity of $\sim7\times10^{40}$~erg~s$^{-1}$, we find that the nominal proton energy density of several eV~cm$^{-3}$ cannot explain the Hitomi result, and a much broader region needs to be illuminated by MeV protons with much higher energy density. However, we need further limitations of the geometry of proton injection and molecular clouds to more accurately estimate the proton-induced Fe-K$\alpha$ flux. 

%\bibitem[Nobukawa et al.(2015)]{2015ApJ...807L..10N} Nobukawa, K.~K., Nobukawa, M., Uchiyama, H., et al.\ 2015, \apjl, 807, L10 
%\bibitem[Kino et al.(2016)]{2016AN....337...47K} Kino, M., Ito, H., Kawakatu, N., et al.\ 2016, Astronomische Nachrichten, 337, 47

%--------------------------------------------------
\subsection{Comparison with other types of AGN}\label{S4-3}
%--------------------------------------------------

The origin of the Fe-K$\alpha$ line of NGC~1275 in 2016, possibly identified to be the low-covering fraction and/or low-column density torus, or the circumnuclear molecular disk, is clearly distinct from those in normal Seyfert galaxies claimed to be located at a region inside the dusty torus (e.g., Shu et al. 2010, 2011). This may reflect some distinction of BLR and dusty torus structures between NGC~1275 and Seyfert galaxies.  A BLR should be present in 2016, because NGC 1275 was at comparable flux in the Hitomi observations to that of the early 1980s when the broad H$\alpha$ line was detected \citep{1997ApJS..112..391H}, and contemporaneous HST optical data derived on 2016 January 7 show an composite H$\alpha$ line whose profile is similar to that reported by \citet{1997ApJS..112..391H}. Thus, we expect a 
BLR FWHM of $\sim 2750$~km~s$^{-1}$ in 2016, which rules out a large contribution of the BLR to the much narrower Hitomi Fe-K$\alpha$ line. 

Although we do not have other information relevant to if a torus was present during the Hitomi observations, the Eddington ratio might drive the structure of the dusty torus as inferred by \citet{2016ApJ...831...37K}. When an AGN is less luminous than the critical luminosity, its dusty torus might not be able to sufficiently inflate and cover a large enough solid angle to produce a high EW Fe-K$\alpha$ line. Presumably related to this, \citet{2017NatAs...1..679R} and \citet{2017Natur.549..488R} showed that the fraction of the obscured AGN decreases with $L_B \lesssim 10^{42}$~erg~s$^{-1}$ and $L_B/L_{\rm Edd} \lesssim 10^{-4}$, respectively. Our results show that NGC~1275 follows this pattern like other low-luminosity AGN, and thus, its Fe-K$\alpha$ line profile might be different from those from normal Seyfert galaxies.

%\bibitem[Kawamuro et al.(2016)]{2016ApJ...831...37K} Kawamuro, T., Ueda, Y., Tazaki, F., Terashima, Y., \& Mushotzky, R.\ 2016, \apj, 831, 37 

Another possibility is that AGN emission line structures are different in radio-loud and radio-quiet AGN even when their luminosities are similar. This picture is consistent with the fact that the amount of reflection, strength and width of the Fe-K$\alpha$ lines are different in radio-loud AGN (e.g., Tazaki et al. 2011, 2013). Because of the lower EW of Fe-K$\alpha$ lines and lower X-ray fluxes, the study of Fe-K$\alpha$ line emission in radio-loud AGN at high energy resolution with the Chandra gratings has been difficult except for a few brightest sources (e.g., \cite{2016ApJ...830...98T}). Thus, X-ray microcalorimeters onboard future satellites, such as the X-ray Astronomical Recovery Mission (XARM) and the Athena X-ray observatory, are essential for further works in this field as shown by the ability of the SXS onboard Hitomi to detect the weak emission line with EW $\sim9$~eV in NGC~1275, the 50th brightest AGN.  
 
%\bibitem[Sambruna et al 2009 Astrophys.J.700:1473-1487,2009) However, Tazaki et al. (2011, 2013) found no significant difference in structures of their dusty tori from those in radio-quiet sources. 

%\bibitem[Koratkar \& Blaes(1999)]{1999PASP..111....1K} Koratkar, A., \& Blaes, O.\ 1999, \pasp, 111, 1 
%\bibitem[Tazaki et al.(2011)]{2011ApJ...738...70T} Tazaki, F., Ueda, Y., Terashima, Y., \& Mushotzky, R.~F.\ 2011, \apj, 738, 70
%\bibitem[Tazaki et al.(2013)]{2013ApJ...772...38T} Tazaki, F., Ueda, Y., Terashima, Y., Mushotzky, R.~F., \& Tombesi, F.\ 2013, \apj, 772, 38
%\bibitem[Tombesi et al.(2016)]{2016ApJ...830...98T} Tombesi, F., Reeves, J.~N., Kallman, T., et al.\ 2016, \apj, 830, 98

%--------------------------------------------------
\section{Conclusion}
%--------------------------------------------------

The SXS onboard the Hitomi satellite detected and resolved the fluorescence Fe-K$\alpha$ line emission from the radio galaxy NGC~1275 at the center of the Perseus cluster of galaxies. This is the first observation of the Fe-K$\alpha$ line from an AGN with the unprecedentedly high energy resolution at $\sim6.4$~keV provided by an X- ray microcalorimeter. The well-constrained line velocity width of $500$--$1600$~km~s$^{-1}$ allows us to rule out the possibility that the Fe-K$\alpha$ line originates from an accretion disk and a BLR. Moreover, analyzing the Chandra and XMM-Newton archival data, we can limit the outer boundary of the Fe-K$\alpha$ source region to be within $\sim1.6$~kpc from the central engine. Taking into account the small EW of the Fe-K$\alpha$ line $\sim 20$~eV against the AGN continuum, the lack of time variability of the line, and its low physical width, the origin of the Fe-K$\alpha$ line is likely to be a low-covering fraction and/or low-column density molecular torus or a circumnuclear molecular disk associated with infrared [Fe II] lines, unlike normal Seyfert galaxies. The present result demonstrates power of the X-ray microcalorimeter spectroscopy of AGN Fe-K$\alpha$ lines, even if lines are very weak with EW $\sim 9$~eV. Future XARM and Athena observations are expected to  bring us many discoveries and important results in the AGN science.

\begin{trueauthors}
H.~Noda led this study, analyzed the SXS data, and wrote the manuscript, along with Y.~Fukazawa, R.~F.~Mushotzky and C.~S.~Reynolds. H.~Noda and F.~S.~Porter contributed to the SXS hardware development including designs, integrations, performance tests, launch campaign, and in-orbit operation. F.~S.~Porter and S.~Nakashima contributed to the in-orbit calibration of the SXS. Y.~Fukazawa performed the SXS pixel map analyses and the PSF-photometry method. K.~Hagino worked for the Monte Carlo simulation and its interpretation. T.~Kawamuro  performed the analyses of the Chandra and XMM-Newton archival data. H.~Noda, Y.~Fukazawa, R.~F.~Mushotzky, C.~S.~Reynolds, T.~Tanaka, T.~Kawamuro, L.~W.~Brenneman, A.~C.~Fabian, M.~Ohno, and C.~Done contributed to the discussion and interpretation parts. L.~C.~Gallo, C.~Pinto, M.~Loewenstein, P.~Gandhi, T.~Yaqoob, M.~Guainazzi, M.~Tsujimoto, and N.~Yamasaki helped to improve the manuscript. 
\end{trueauthors}

\begin{ack}

We thank the support from the JSPS Core-to-Core Program.
We acknowledge all the JAXA members who have contributed to the ASTRO-H (Hitomi)
project.
All U.S. members gratefully acknowledge support through the NASA Science Mission
Directorate. Stanford and SLAC members acknowledge support via DoE contract to SLAC
National Accelerator Laboratory DE-AC3-76SF00515. Part of this work was performed under
the auspices of the U.S. DoE by LLNL under Contract DE-AC52-07NA27344.
Support from the European Space Agency is gratefully acknowledged.
French members acknowledge support from CNES, the Centre National d'\'{E}tudes Spatiales.
SRON is supported by NWO, the Netherlands Organization for Scientific Research.  Swiss
team acknowledges support of the Swiss Secretariat for Education, Research and
Innovation (SERI).
The Canadian Space Agency is acknowledged for the support of Canadian members.  
We acknowledge support from JSPS/MEXT KAKENHI grant numbers 15J02737,
15H00773, 15H00785, 15H02090, 15H03639, 15H05438, 15K05107, 15K17610,
15K17657, 16J00548, 16J02333, 16H00949, 16H06342, 16K05295, 16K05296,
16K05300, 16K13787, 16K17672, 16K17673, 21659292, 23340055, 23340071,
23540280, 24105007, 24244014, 24540232, 25105516, 25109004, 25247028,
25287042, 25400236, 25800119, 26109506, 26220703, 26400228, 26610047,
26800102, JP15H02070, JP15H03641, JP15H03642, JP15H06896,
JP16H03983, JP16K05296, JP16K05309, JP16K17667, and JP16K05296.
The following NASA grants are acknowledged: NNX15AC76G, NNX15AE16G, NNX15AK71G,
NNX15AU54G, NNX15AW94G, and NNG15PP48P to Eureka Scientific.
H. Akamatsu acknowledges support of NWO via Veni grant.  
C. Done acknowledges STFC funding under grant ST/L00075X/1.  
A. Fabian and C. Pinto acknowledge ERC Advanced Grant 340442.
P. Gandhi acknowledges JAXA International Top Young Fellowship and UK Science and
Technology Funding Council (STFC) grant ST/J003697/2. 
Y. Ichinohe, K. Nobukawa, and H. Seta are supported by the Research Fellow of JSPS for Young
Scientists.
N. Kawai is supported by the Grant-in-Aid for Scientific Research on Innovative Areas
``New Developments in Astrophysics Through Multi-Messenger Observations of Gravitational
Wave Sources''.
S. Kitamoto is partially supported by the MEXT Supported Program for the Strategic
Research Foundation at Private Universities, 2014-2018.
B. McNamara and S. Safi-Harb acknowledge support from NSERC.
T. Dotani, T. Takahashi, T. Tamagawa, M. Tsujimoto and Y. Uchiyama acknowledge support
from the Grant-in-Aid for Scientific Research on Innovative Areas ``Nuclear Matter in
Neutron Stars Investigated by Experiments and Astronomical Observations''.
N. Werner is supported by the Lend\"ulet LP2016-11 grant from the Hungarian Academy of
Sciences.
D. Wilkins is supported by NASA through Einstein Fellowship grant number PF6-170160,
awarded by the Chandra X-ray Center, operated by the Smithsonian Astrophysical
Observatory for NASA under contract NAS8-03060.

We thank contributions by many companies, including in particular, NEC, Mitsubishi Heavy
Industries, Sumitomo Heavy Industries, and Japan Aviation Electronics Industry. Finally,
we acknowledge strong support from the following engineers.  JAXA/ISAS: Chris Baluta,
Nobutaka Bando, Atsushi Harayama, Kazuyuki Hirose, Kosei Ishimura, Naoko Iwata, Taro
Kawano, Shigeo Kawasaki, Kenji Minesugi, Chikara Natsukari, Hiroyuki Ogawa, Mina Ogawa,
Masayuki Ohta, Tsuyoshi Okazaki, Shin-ichiro Sakai, Yasuko Shibano, Maki Shida, Takanobu
Shimada, Atsushi Wada, Takahiro Yamada; JAXA/TKSC: Atsushi Okamoto, Yoichi Sato, Keisuke
Shinozaki, Hiroyuki Sugita; Chubu U: Yoshiharu Namba; Ehime U: Keiji Ogi; Kochi U of
Technology: Tatsuro Kosaka; Miyazaki U: Yusuke Nishioka; Nagoya U: Housei Nagano;
NASA/GSFC: Thomas Bialas, Kevin Boyce, Edgar Canavan, Michael DiPirro, Mark Kimball,
Candace Masters, Daniel Mcguinness, Joseph Miko, Theodore Muench, James Pontius, Peter
Shirron, Cynthia Simmons, Gary Sneiderman, Tomomi Watanabe; ADNET Systems: Michael
Witthoeft, Kristin Rutkowski, Robert S. Hill, Joseph Eggen; Wyle Information Systems:
Andrew Sargent, Michael Dutka; Noqsi Aerospace Ltd: John Doty; Stanford U/KIPAC: Makoto
Asai, Kirk Gilmore; ESA (Netherlands): Chris Jewell; SRON: Daniel Haas, Martin Frericks,
Philippe Laubert, Paul Lowes; U of Geneva: Philipp Azzarello; CSA: Alex Koujelev, Franco
Moroso.

\end{ack}

%--------------------------Appendix------------------------
\appendix
\section{The SXS PSF-photometry analyses}
%--------------------------Appendix------------------------

\begin{figure}[t]
\FigureFile(75mm,75mm){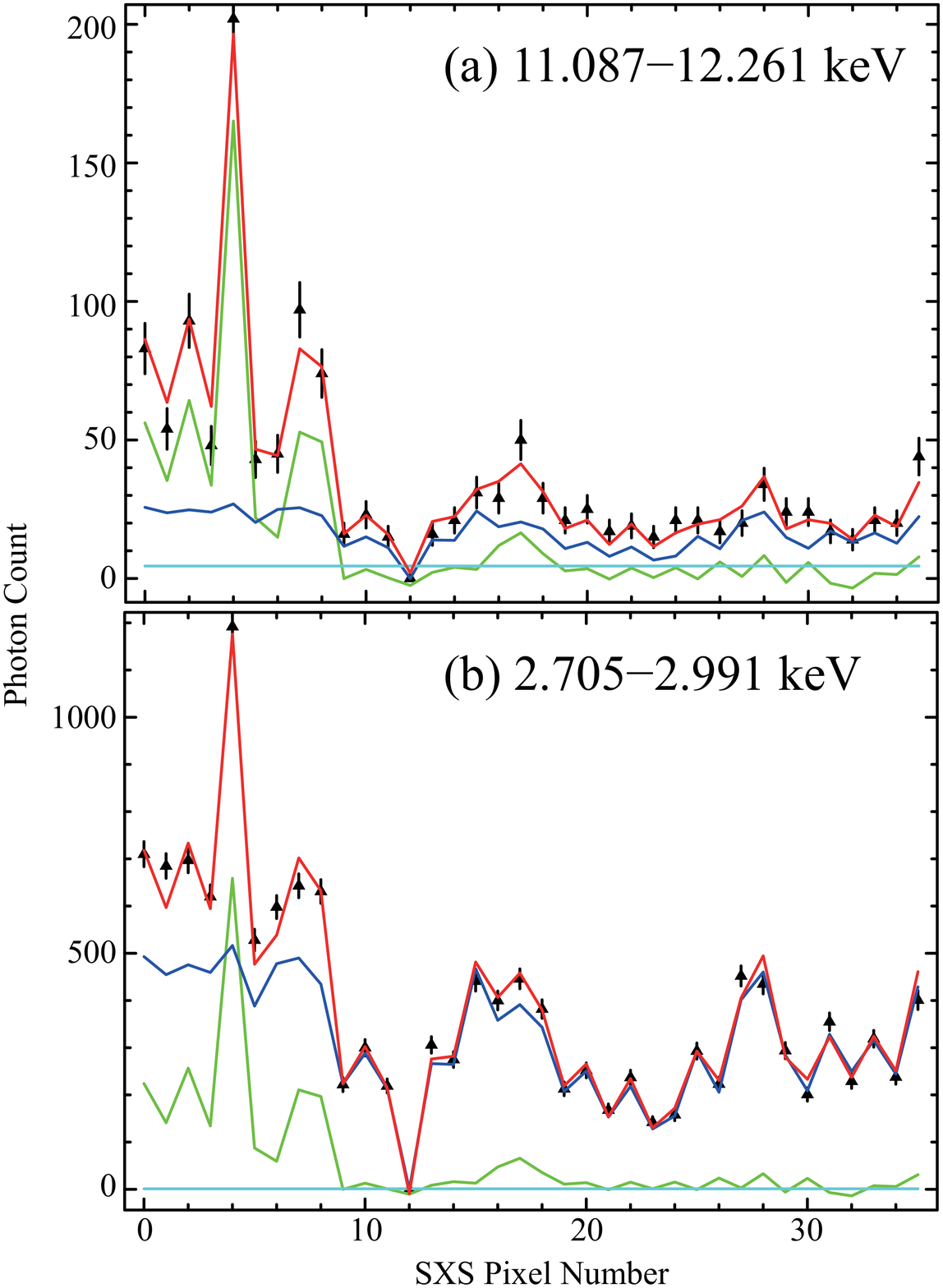}
\vspace{0.3cm}
\caption{The SXS image fitting results of the 11.087--12.261~keV band (panel a) and the 2.705--2.991~keV band (panel b) in the PSF photometry method. Triangle data show a photon count in each SXS pixel. Blue, green, and cyan represent a model component of ICM, AGN, and background, respectively, while red shows a total model. }
\label{fig:psf-photometry}
\end{figure}

When we apply the PSF-photometry analyses to the SXS pixel map, we require that the satellite attitude is stable, and hence, we utilize the same GTI as used in \S3.1.2. In detail, we create an SXS count map in 20 energy bands, where energy boundaries are logarithmically spaced in 2--15 keV. Because photon statistics in the higher energy bands are limited, we use wider energy bands above15~keV; 15--17~keV and 17--20~keV.  In addition, we apply finer energy bands around the Fe-K$\alpha$ line and Fe-K edge, avoiding the Fe~XXV~He$\alpha$ and Fe~XXVI~H$\alpha$ line complex; 6.0575--6.25, 6.25--6.26, ...(0.01~keV step) , 6.32--6.33, 6.33--6.47, 6.62--6.78, 6.90--6.92, ...(0.02~keV step), 7.06--7.08, and 7.08--7.41~keV.  We then fit each count map with a spatial model consisting of ICM + AGN + background as shown in figure~\ref{fig:psf-photometry}. The spatial distribution of the ICM in each band is assumed to follow that of the Fe~XXV~He$\alpha$ line complex (6.52--6.59~keV), with an iterative correction applied to account for the (small) contribution of the AGN to this band. Background counts are estimated from the public SXS background data, and then, it is assumed to be uniform over all the SXS pixels. 

%%%%%%%%%%%%%%%%figure8%%%%%%%%%%%%%%%%%%%%%%%%
%%%%%%%%%%%%%%%%figure8%%%%%%%%%%%%%%%%%%%%%%%%

In the estimation of the AGN emission spatial distribution, we create a spatial model by using the count map in 10--12~keV, and subtracting the best-fit ICM emission and background model from the source image. In the count map fits, the AGN spatial model is not complete because of the PSF systematic errors and the attitude fluctuation, and hence, the count maps are not well reproduced around the AGN. Therefore, we include the systematic errors to the estimated counts derived by the PSF photometry, so that the reduced $\chi^2$ value of the count map fitting becomes 1. After fitting the SXS pixel map in each energy band, we calculate the model count rate of AGN in each energy bin in the center on-core 9 pixels, and their errors are propagated from the image fitting errors. As a result, we derive a PSF-photometry spectrum of the AGN shown in figure~\ref{fig:SXSbroadband} red.

%%%%%%%%%%%%%%%%figure5%%%%%%%%%%%%%%%%%%%%%%%%
\begin{figure}[t]
\begin{center}
\FigureFile(75mm,75mm){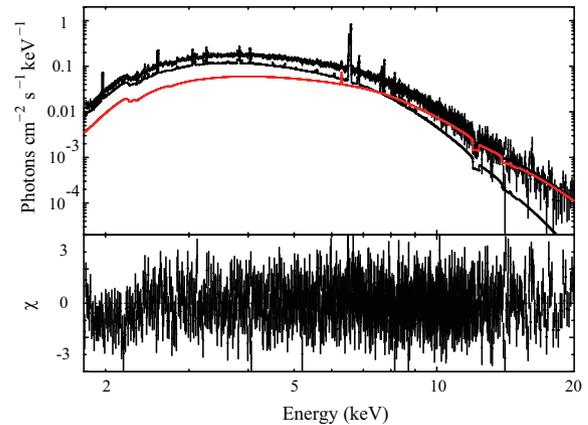}
\end{center}
\caption{Same as panel (a) in figure~\ref{fig:SXSbroadband} except for the absence of the PSF-photometry spectrum, but the AGN fitting parameters are limited within 90\% errors which are derived in the fit to the PSF-photometry spectrum. }
\label{fig:SXSbroadband_limited}
\end{figure}
%%%%%%%%%%%%%%%%figure5%%%%%%%%%%%%%%%%%%%%%%%%

Because the AGN continuum shape looks different between the direct fit to the SXS on-core 9-pixels spectrum and the PSF-photometry spectrum fit as shown in figure~\ref{fig:SXSbroadband}(a), the consistency between them should be checked. Then, we fit the SXS on-core 9-pixels spectrum with the same model as that in \S3.2.1, but with the {\tt pegpwrlw} photon index and the 2--10 keV flux limited to be free within the 90\% error ranges derived by the PSF-photometry spectrum fit (the Baseline column in table~\ref{tab:SXSbroadband}). Figure~\ref{fig:SXSbroadband_limited} shows the result, and the ICM emission parameters do not significantly change from those without the AGN parameter limitations (\S3.2.1), as $z_{\rm ICM}= 0.01729^{+0.00002}_{-0.00004}$, $T_{\rm e}=3.80 \pm 0.06$~keV, and $V_{\rm turb}=157.8 \pm 11.8$~km~s$^{-1}$. The fit gives $C$-statistics/d.o.f.$=10568.6/11120$, and its difference ($\Delta C$-statistics) from the direct fit is only $6.4$, which is statistically allowed. Thus, the PSF-photometry spectrum is confirmed to be consistent with the AGN spectral shape derived by the direct fit in \S3.2.1.

%Because we confirm that the AGN emission dominates above $\sim10$~keV, we check the light curve of NGC~1275 by extracting the data of the  SXS center  $3\times3$ pixels in 10--15~keV finding no significant count rate change within 15\% (root-mean-square)  during the SXS observation from February 25--27 and March 4--6 in 2016.  

%--------------------------Appendix------------------------
\section{Monte-Carlo Simulation}
%--------------------------Appendix------------------------

Usually, the fluorescence narrow and neutral Fe-K$\alpha$ line of AGN is thought to come from a BLR and/or a molecular torus. Estimation of such a Fe-K$\alpha$ line intensity has been reported by various authors (e.g. \cite{2009ApJ...692..608I}; \cite{2009MNRAS.397.1549M}; \cite{2014ApJ...787...52L}; \cite{2016ApJ...818..164F}). However, the central dominant galaxy in the cluster of galaxies like NGC~1275 has other candidates of the fluorescence Fe-K$\alpha$ line source, as suggested by \citet{Chur98}. Around NGC~1275, prominent H$\alpha$ filamentary structure and CO molecular clouds were reported (e.g., \cite{2006A&A...454..437S}).  When such CO clouds are illuminated by the AGN or the ICM X-ray continua, fluorescence Fe-K$\alpha$ line can be generated. Then, we employ the Monte-Carlo simulation to estimate Fe-K$\alpha$ flux from the molecular clouds by utilizing the {\tt MONACO} framework (\cite{2011ApJ...740..103O}; \cite{2015MNRAS.446..663H}). Mass of each molecular cloud is set to be $10^9\exp{\left(-r/14\right)} M_{\odot}$ against the distance $r$~kpc from the NGC~1275 center, and the hydrogen number density is assumed to $10^3$~cm$^{-3}$, which means that the typical cloud radius is $\simeq0.2$~kpc. Radial distribution of clouds is calculated by differentiating the accumulated mass profile of $4.02\times10^9 \log{\left(0.91r\right)} M_{\odot}$ with 5~kpc grids, and the clouds are placed spherically symmetric with a random angular distribution. We assumed an intracloud velocity dispersion to be $300$--$8r$ km s$^{-1}$, and a bulk cloud velocity as 200~km~s$^{-1}$ in random directions. All of the physical properties defined above are based on the observed cloud properties reported in \citet{2006A&A...454..437S}.

%%%%%%%%%%%%%%%%figure8%%%%%%%%%%%%%%%%%%%%%%%%
\begin{figure}[t]
\FigureFile(80mm,80mm){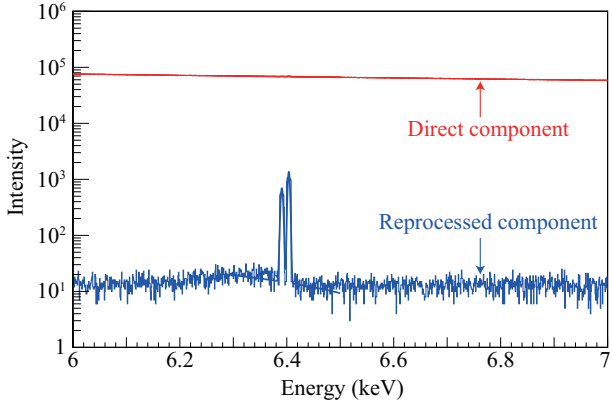}
\vspace{0.3cm}
\caption{The Monte-Carlo simulation result by using MONACO. Red shows a direct X-ray continuum, while blue represents a reprocessed emission including Fe-K$\alpha_1$ and K$\alpha_2$ lines.  }
\label{fig:monaco}
\end{figure}
%%%%%%%%%%%%%%%%figure8%%%%%%%%%%%%%%%%%%%%%%%%

The AGN input continuum spectrum is assumed to follow a \texttt{powerlaw} with a photon index of 1.7. ICM emission spectrum is assumed to follow a bremsstrahlung emission with a temperature of 3.8~keV, and radial intensity distribution to follow a $\beta$ model with a core radius of 26~kpc and a $\beta$ of 0.53 (\cite{2015MNRAS.450.4184Z}). As shown in figure~\ref{fig:monaco}, the EW of Fe-K$\alpha$ line against the input continuum spectrum becomes 0.15~eV and 0.004~eV for illuminating source of AGN and ICM, respectively. Thus, the Monte-Carlo simulation with the MONACO framework revealed that the molecular clouds outside the NGC~1275 generate too small Fe-K$\alpha$ flux to explain EW$\sim 20$~eV obtained by Hitomi/SXS, even when we consider an entire region of the 15~kpc filament.

%\bibitem[Salom{\'e} et al.(2006)]{2006A&A...454..437S} Salom{\'e}, P., Combes, F., Edge, A.~C., et al.\ 2006, \aap, 454, 437 
%\bibitem[Odaka et al.(2011)]{2011ApJ...740..103O} Odaka, H., Aharonian, F., Watanabe, S., et al.\ 2011, \apj, 740, 103  

%\bibitem[Churazov et al.(1998)]{Chur98} Churazov, E., Sunyaev, R., Gilfanov, M., Forman, W., \& Jones, C.\ 1998, \mnras, 297, 1274 
%\bibitem[Ikeda et al.(2009)]{2009ApJ...692..608I} Ikeda, S., Awaki, H., \& Terashima, Y.\ 2009, \apj, 692, 608 
%\bibitem[Liu \& Li(2014)]{2014ApJ...787...52L} Liu, Y., \& Li, X.\ 2014, \apj, 787, 52
%\bibitem[Murphy \& Yaqoob(2009)]{2009MNRAS.397.1549M} Murphy, K.~D., \& Yaqoob, T.\ 2009, \mnras, 397, 1549 
%\bibitem[Furui et al.(2016)]{2016ApJ...818..164F} Furui, S., Fukazawa, Y., Odaka, H., et al.\ 2016, \apj, 818, 164 
%\bibitem[Hagino et al.(2015)]{2015MNRAS.446..663H} Hagino, K., Odaka, H., Done, C., et al.\ 2015, \mnras, 446, 663 
%\bibitem[Zhuravleva et al.(2015)]{2015MNRAS.450.4184Z} Zhuravleva, I., Churazov, E., Ar{\'e}valo, P., et al.\ 2015, \mnras, 450, 4184 

\end{document}